\documentclass[journal,comsoc]{IEEEtran}
\usepackage[ruled]{algorithm2e}

\SetAlFnt{\small}
\SetAlCapFnt{\small}
\SetAlCapNameFnt{\small}
\SetAlCapHSkip{0pt}
\IncMargin{-\parindent}

\usepackage{framed,color}
\usepackage{xparse}
\usepackage{nccmath}
\usepackage[utf8]{inputenc}
\usepackage[T1]{fontenc}
\usepackage{amsmath,amssymb,array,color,colortbl,graphicx,multirow}
\usepackage{microtype}
\usepackage{comment}
\usepackage[caption=false]{subfig} 
\usepackage{times}
\usepackage{here}
\usepackage{multirow} 
\usepackage{etoolbox}
\usepackage[normalem]{ulem}
\usepackage{enumitem}

\usepackage{makecell}

\usepackage{url}

\newtoggle{comments}
\toggletrue{comments} 

\iftoggle{comments}{	\newcommand{\klaus}[1]{\textit{\textcolor{red}{[klaus]: #1}}} 
	\newcommand{\stefano}[1]{\textit{\textcolor{brown}{[stefano]: #1}}} 
	\newcommand{\stefan}[1]{\textit{\textcolor{blue}{[stefan]: #1}}} 
}{	\newcommand{\klaus}[1]{} 
	\newcommand{\stefano}[1]{} 
	\newcommand{\stefan}[1]{} 
}

\newcommand{\rev}[1]{#1}

\usepackage{tikz}
\usetikzlibrary{positioning,shapes,shadows,arrows,chains,decorations.pathreplacing,decorations.pathmorphing, shapes.arrows, fadings, calc, decorations.markings, shapes.geometric, automata, er, matrix, decorations.markings}

\tikzset{>=latex} 

\newcommand{\congestion}{capacity}
\newcommand{\Congestion}{Capacity}

\usepackage{pgfplots}

\pgfdeclarelayer{overlay}
\pgfdeclarelayer{underlay}
\pgfdeclarelayer{edgelayer}
\pgfdeclarelayer{nodelayer}
\pgfsetlayers{underlay,main,overlay,edgelayer,nodelayer}

\pgfplotsset{compat=1.14}

\tikzset{markovstate/.style={shape=circle,draw=black,align=center,inner sep=0,minimum width=7mm}}
\tikzset{markovedge/.style={-latex}}
\tikzset{label/.style={}}
\definecolor{color1}{RGB}{0,200,100}
\definecolor{color2}{RGB}{200,100,0}
\definecolor{color3}{RGB}{100,0,200}
\definecolor{color4}{RGB}{200,0,0}

\tikzstyle{none}=[inner sep=0pt]
\definecolor{hexcolor0xffffff}{rgb}{1.000,1.000,1.000}
\definecolor{hexcolor0x000000}{rgb}{0.000,0.000,0.000}
\definecolor{hexcolor0x00ff00}{rgb}{0.000,1.000,0.000}
\definecolor{hexcolor0x000000}{rgb}{0.000,0.000,0.000}
\definecolor{hexcolor0xffff00}{rgb}{1.000,1.000,0.000}
\definecolor{hexcolor0x000000}{rgb}{0.000,0.000,0.000}
\definecolor{hexcolor0xffffff}{rgb}{1.000,1.000,1.000}
\definecolor{hexcolor0x000000}{rgb}{0.000,0.000,0.000}
\definecolor{hexcolor0x000000}{rgb}{0.000,0.000,0.000}
\definecolor{hexcolor0x000000}{rgb}{0.000,0.000,0.000}

\tikzstyle{rn}=[circle,fill=hexcolor0xffffff,draw=hexcolor0x000000,line width=0.8 pt]
\tikzstyle{gn}=[circle,fill=hexcolor0x00ff00,draw=hexcolor0x000000,line width=0.8 pt]
\tikzstyle{yn}=[circle,fill=hexcolor0xffff00,draw=hexcolor0x000000,line width=0.8 pt]
\tikzstyle{newstyle}=[circle,fill=hexcolor0xffffff,draw=hexcolor0x000000]

\tikzstyle{simple}=[-,draw=hexcolor0x000000,line width=2.000]
\tikzstyle{arrow}=[-,draw=hexcolor0x000000,postaction={decorate},decoration={markings,mark=at position .5 with {\arrow{>}}},line width=2.000]
\tikzstyle{tick}=[-,draw=hexcolor0x000000,postaction={decorate},decoration={markings,mark=at position .5 with {\draw (0,-0.1) -- (0,0.1);}},line width=2.000]
\tikzstyle{newstyle}=[-latex,draw=hexcolor0x000000]
\tikzstyle{dashed}=[-latex,draw=hexcolor0x000000]
\tikzstyle{rect}=[rectangle,fill=hexcolor0xffffff,draw=hexcolor0x000000]
\tikzstyle{elli}=[ellipse,fill=hexcolor0xffffff,draw=hexcolor0x000000]

\newcommand{\etal}{\textit{et al}. }

\usepackage{balance}

\RequirePackage{color} \RequirePackage[normalem]{ulem} \newcommand\hsout[1]{\let\helpcmd\sout\parhelp#1\par\relax\relax}   \long\def\parhelp#1\par#2\relax{    \helpcmd{#1}\ifx\relax#2\else\par\parhelp#2\relax\fi  }

\begin{document}
\bstctlcite{MyBSTcontrol}

\makeatletter
\newcommand{\removelatexerror}{\let\@latex@error\@gobble}
\makeatother

\def\NULL/{\textnormal{\texttt{null}}}

\newcounter{ipCounter}
\NewDocumentEnvironment{IPFormulation}{m}{\refstepcounter{ipCounter}
\begin{algorithm}[#1]\renewcommand\thealgocf{\arabic{ipCounter}}
}{\end{algorithm}
\addtocounter{algocf}{-1}
}

\newcommand{\tagIt}{\refstepcounter{IPnumber}\tag{IP-\theIPnumber}}
\newcounter{IPnumber}
\setcounter{IPnumber}{0}

\newenvironment{ORIG}{\par\color{blue} \textcolor{red}{[BEGIN Vassilis' original]}}{\textcolor{red}{[END Vassilis' original]} \par}

\newenvironment{REWRITTEN}{\par\color{blue} \textcolor{red}{[BEGIN Vassilis' original: deprecated]}}{\textcolor{red}{[END Vassilis' original: deprecated]} \par}

	\newcommand{\deltaOut}[1]{\delta^+_{#1}}
	\newcommand{\deltaIn}[1]{\delta^-_{#1}}
	\newcommand{\head}{\textnormal{head}}
	\newcommand{\tail}{\textnormal{tail}}
	\newcommand{\negate}[1]{\overline{#1}}

	\newcommand{\fontMacro}[1]{\mathsf{#1}}
	\newcommand{\fontVariable}[1]{\mathrm{#1}}
	\newcommand{\fontParameter}[1]{\mathbf{#1}}
	\newcommand{\fontParameterElement}[1]{\textnormal{#1}}
	\newcommand{\fontConstraintName}[1]{\textsc{#1}}
	\newcommand{\foM}[1]{\fontMacro{#1}}
	\newcommand{\foV}[1]{\fontVariable{#1}}
	\newcommand{\foP}[1]{\fontParameter{#1}}
	\newcommand{\foPE}[1]{\fontParameterElement{#1}}
	\newcommand{\foCN}[1]{\fontConstraintName{#1}}

\newcommand{\nat}{\mathbb{N}}
\newcommand{\psnat}{\mathbb{N}_{> 0}}
\newcommand{\reals}{\mathbb{R}}
\newcommand{\preals}{\mathbb{R}_{\geq 0}}
\newcommand{\nreals}{\mathbb{R}_{\leq 0}}

\newcommand{\psreals}{\mathbb{R}_{> 0}}
\newcommand{\nsreals}{\mathbb{R}_{< 0}}

\newcommand{\integers}{\mathbb{Z}}
\newcommand{\pint}{\mathbb{Z}_{\geq 0}}
\newcommand{\nint}{\mathbb{Z}_{\leq 0}}

\newcommand{\psint}{\mathbb{Z}_{> 0}}
\newcommand{\nsint}{\mathbb{Z}_{< 0}}

\newcommand{\V}{V}
\newcommand{\EO}{E_{\pi_1}}
\newcommand{\EN}{E_{\pi_2}}
\newcommand{\EWWP}{E_{\overline{\textsf{WP}}}}
\newcommand{\EG}{E}
\newcommand{\lat}{lat}
\newcommand{\START}{s}
\newcommand{\END}{t}
\newcommand{\W}{wp}
\newcommand{\FP}{\mathcal{P}_{\textnormal{forbidden}}}
\newcommand{\PC}{\mathcal{P}_E^C}
\newcommand{\PWP}{\mathcal{P}^{\overline{WP}}_E}

\newcommand{\R}{\mathcal{R}}
\newcommand{\PreR}{\mathcal{R}_p}
\newcommand{\SupR}{\mathcal{R}_0}

\newcommand{\upgradeNode}{x}
\newcommand{\edgeExists}{y}
\newcommand{\edgeExistsTransient}{y}
\newcommand{\nodeReachable}{a}
\newcommand{\nodeReachableWOWP}{\overline{a}}
\newcommand{\nodeLevel}{l}
\newcommand{\distanceToEnd}{dist}
\newcommand{\flow}{f}
\newcommand{\roundTime}{t}
\newcommand{\updateInRound}{U}
\newcommand{\numberOfUsedRounds}{R}
\newcommand{\VWP}{\ensuremath{\mathit{WP}}}

\markboth{K.-T. Foerster, S. Schmid, and S. Vissicchio}{Survey of Consistent {Software-Defined} Network Updates}

\title{Survey of Consistent\\ Software-Defined Network Updates}

\author{Klaus-Tycho Foerster,
        Stefan Schmid,
        and~Stefano Vissicchio				\\ \vspace{3mm}
				\small \copyright 2018 IEEE. This is the author's version of an article that will appear in IEEE Communications Surveys \& Tutorials.\\ The final version of record is available at \url{http://dx.doi.org/10.1109/COMST.2018.2876749}.
\vspace{-6mm}\thanks{K.-T. Foerster and S. Schmid are with University of Vienna, Vienna, Austria. E-mail: klaus-tycho.foerster@univie.ac.at and stefan\_schmid@univie.ac.at.}\thanks{S. Vissicchio is with University College London, London, United Kingom. E-mail: s.vissicchio@cs.ucl.ac.uk. }\thanks{Manuscript sent February '18, revised August '18, accepted October '18.}}

\sloppy

\maketitle

\begin{abstract}
Computer networks have become a critical infrastructure.
In fact, networks should not only meet strict requirements in terms of correctness, availability, and performance, but they should also be very flexible and support fast updates, e.g., due to policy changes, increasing traffic, or failures.
This paper presents a structured survey of mechanism and protocols to update computer networks in a fast and consistent manner. 
In particular, we identify and discuss the different desirable consistency properties that should be provided throughout a network update, the algorithmic techniques which are needed to meet these consistency properties, and the implications on the speed and costs at which updates can be performed. 
We also explain the relationship between consistent network update problems and classic algorithmic optimization ones. 
While our survey is mainly motivated by the advent of Software-Defined Networks (SDNs) and their primary need for correct and efficient update techniques, the fundamental underlying problems are not new, and we provide a historical perspective of the subject as well.
\end{abstract}

\section{Introduction}\label{sec:intro}

Computer networks such as datacenter networks,
enterprise networks, carrier networks etc.~have become
a critical infrastructure of the information society.
The importance of computer networks
and the resulting strict requirements in terms of
availability, performance, and correctness  
however stand in contrast to today's ossified
computer networks: the techniques and 
methodologies used to build, manage, and debug 
computer networks are largely the same as those used in 1996~\cite{barefoot}.
Indeed, operating traditional computer networks is often a cumbersome
and error-prone task, and even tech-savvy
companies such as GitHub, Amazon, GoDaddy, etc.~frequently 
report issues with their network, due to
misconfigurations, e.g., resulting in forwarding loops~\cite{z1,z2,z3,z4}.
An anecdote reported in~\cite{barefoot} illustrating the problem, is the
one by a Wall Street investment bank: due to a datacenter outage, the bank was
suddenly losing millions of dollars per minute. Quickly the compute and storage
emergency teams compiled a wealth of information giving insights into what might
have happened. In contrast, the networking team only had very primitive
connectivity testing tools such as ping and traceroute, to debug the problem.
They could not provide any insights into the actual problems of the switches or
the congestion experienced by individual packets, nor could the team create any
meaningful experiments to identify, quarantine and resolve the
problem~\cite{barefoot}.

Software-defined networking is an interesting new paradigm
which allows to operate and verify networks in a more principled
and formal manner, while also introducing flexibilities
and programmability, and hence foster innovations. 
In a nutshell, a Software-Defined Network (SDN)
outsources and consolidates the control over the forwarding 
or routing devices
(located in the so-called \emph{data plane})  to a logically
centralized controller software (located in the 
so-called \emph{control plane}).
This decoupling allows to evolve and innovate
the control plane independently from the hardware
constraints
of the data plane. Moreover, OpenFlow, the \rev{{\emph{de facto}}}
standard SDN protocol today, is based on a simple
match-action paradigm: the behavior of
an OpenFlow switch is defined by a set of
forwarding rules installed by the controller.
Each rule consists of a match and an
action part: all packets matched by a given rule
are subject to the corresponding action.
Matches are defined over Layer-2 to Layer-4
header fields (e.g., MAC and IP addresses, TCP ports, etc.),
and actions typically describe operations such as
forward to a specific port, drop, or update certain header fields.
In other words, in an SDN/OpenFlow network, network devices
become simpler: their behavior is defined by a set of rules installed
by the controller. This enables formal reasoning and verification,
as well as flexible network update,
from a logically centralized perspective~\cite{hsa,veriflow}.
Moreover, as rules can be defined over multiple OSI layers, the distinction
between switches and routers (and even simple middleboxes~\cite{road})
becomes blurry.

However, the decoupling of the control plane from the data plane
also introduces new challenges. In particular, the switches and controllers
as well as their interconnecting network form a complex asynchronous
distributed system. For example, a remote controller may
learn about and react to network events slower (or not at all)
than a hardware device in the
data plane: given a delayed and inconsistent view, a controller
(and accordingly the network)
may behave in an undesirable way. Similarly, new rules
or rule updates communicated from the controller(s) to the 
switch(es) may take effect in a delayed and asynchronous manner:
not only because
these updates have to be transmitted from the controller to the switches
over the network, but also the reaction time of the switches themselves
may differ (depending on the specific hardware, data 
structures, or concurrent load).

Thus, while SDN offers great opportunities to operate
a network in a correct and verifiable manner, there remains
a fundamental
challenge of how to deal with the asynchrony inherent
in the communication channel between controller and switches
as well as in the switches themselves.
Accordingly, the question of how to update network behavior and configurations
correctly yet 
efficiently has been studied intensively over the last years.
However, the notions of correctness and efficiency
significantly differ across the literature. Indeed,
what kind of correctness is needed and which performance
aspects are most critical often depends on the context: 
in security-critical networks, a very strong notion of correctness may be needed, 
even if it comes at a high performance cost;
in other situations, however, short transient inconsistencies
may be acceptable, as long as at least some more basic consistency
guarantees
are provided (e.g., loop-freedom).

We observe that not only is the number of research results in the area growing
very quickly, but also the number of models, the different notions of
consistency and optimization objectives, as well as the algorithmic techniques:
Thus, it has become difficult to keep an overview of the field even for active
researchers. Moreover, many of the underlying update problems are not
entirely new or specific to SDN: rather, techniques to consistently update legacy
networks have been studied in the literature, although they are based on the (more
restrictive) primitives available in traditional protocols (e.g., IGP weights).

We therefore believe that it is time for a comprehensive survey of the
subject.

\subsection{The Network Update Problem}

Any dependable network does not only need to maintain certain invariants, related to correctness, availability, and performance, but also needs to be flexible in how it process~packets. 
\subsubsection{\rev{Flexibility}}
Flexibility implies that networks have to be updated, e.g., to support the following use cases.
\paragraph{Security policy changes}
 For example, in enterprise networks, traffic from a specific subnetwork may have to be routed via a firewall if specific alarms are raised. Similarly, in wide-area networks, the countries that must be avoided by sensitive traffic can change over time.

\paragraph{Traffic engineering}
To improve performance metrics (e.g., minimizing the maximal link load), a network operator may decide to reroute some traffic to different paths. For example, many Internet Service Providers change their paths during the day, depending on the expected load or in reaction to external changes (e.g., a policy modification from a content provider).

\paragraph{Maintenance work}
Also maintenance work may require the update of network routes. For example, in order to replace a faulty router, or to upgrade an existing router, it can be necessary to temporarily reroute traffic.

\paragraph{Link and node failures}
Failures happen quite 
frequently and unexpectedly in today's computer networks, and typically require a fast reaction. Accordingly, fast network monitoring and update mechanisms are required to react to such failures, e.g., by determining a failover path.

\paragraph{Service relocation}
Networks typically run several services, from in-network packet processing functions (e.g., virtualized middleboxes) to applications (like data storage or application servers). Addition, removal or relocation of any of those services would require a network update, i.e., to reroute traffic for the affected service.

\subsubsection{\rev{Maintaining consistency}} It is often desirable that the network maintains certain consistency properties \emph{throughout the update}. Those properties may include per-packet path consistency (a packet should be forwarded along the old or the new route, but never a mixture of both), waypoint enforcement (a packet should never bypass a firewall), or at least correct packet delivery (at no point in time packets should be dropped or trapped in a loop).

\subsubsection{\rev{Towards SDNs}}While the above reasons for network updates are relevant independently
of the adopted paradigm,
the decoupling of control- and data-plane, as well as the flexibility allowed by
the SDN architecture are likely to increase the frequency of network updates,
e.g., for supporting more fine-grained and frequent optimization of traffic
paths~\cite{b4}.

\subsection{Our Contributions}

This paper presents a comprehensive survey of the consistent network update
problem in Software-Defined Networks (SDNs).
\rev{
In the basic scenario assumed by most prior contributions, an SDN network is
controlled by a single controller, which needs to preserve specific consistency
properties at each and every moment during the update.
Preserving such properties is often argued to be more important that the induced
inability to guarantee perfect network availability or partition
tolerance simultaneously---e.g., to avoid packet losses or security
breaches.
This impossibility result follows from the celebrated CAP theorem~\cite{DBLP:journals/sigact/GilbertL02}, which also applies to control algorithms used in networks~\cite{cap-hotsdn13}. 
Throughout this paper, we first consider this basic scenario and then extend the
discussion to network updates with distributed SDN controllers and different
consistency models~\cite{7997164,6133253,DBLP:books/daglib/0019513}.
}
The goal of our survey is to both (1) provide active researchers in the field
with an overview of the state-of-the-art literature, and (2) help researchers
who only recently became interested in the subject bootstrap and learn about
open research questions.

In discussing the literature, we identify and compare the consistency properties
(absence of forwarding loops and blackholes, policy preservation, congestion
avoidance, etc.) and performance objectives (update duration, maximum link
overload during the update, etc.) considered by the scientific literature. We
provide an overview of the algorithmic techniques required to solve specific
classes of network update problems, and discuss the inherent tradeoffs between
the achievable level of consistency and the speed at which networks can be
updated. In fact, as we will see, some update techniques are not only less
efficient than others, but with them, it can even be impossible to consistently
update a network.

We also present a historical perspective, surveying 
the consistency notions provided in traditional
networks and discussing the corresponding techniques
accordingly. 

Moreover, we put the algorithmic
problems into perspective and discuss how these problems
relate to classic optimization and graph theory problems,
such as multi-commodity flow problems or maximum
acyclic subgraph problems.

\subsection{Paper Organization}

The remainder of this paper is organized
as follows. 
\S\ref{sec:history} presents a historical perspective
and reviews notions of consistency and techniques 
both in traditional computer networks as well as
in Software-Defined Networks. 
\S\ref{sec:taxo} 
then presents a classification and taxonomy of the 
different variants of the consistent network update problems.
\S\ref{sec:forwarding}, \S\ref{sec:policies},
and \S\ref{sec:cap} review models and techniques
for connectivity consistency, policy consistency, and \congestion~consistency related problems, respectively.
\S\ref{sec:orthogonal} discusses proposals
to further relax consistency guarantees by introducing
tighter synchronization. 
In \S\ref{sec:practice}, we identify
practical challenges.
After highlighting future research directions
in \S\ref{sec:openprob},
we conclude our paper in \S\ref{sec:conclusion}.

\section{\rev{History of the Network Update Problem\\from the Origins to SDN}}\label{sec:history}

Any computer network must guarantee some consistency properties for the
configured forwarding rules and paths. For example, forwarding loops must be
avoided, as they can quickly deplete switch buffers and harm the availability
and connectivity provided by a network.

It is eminently desirable to preserve consistency properties \textit{during
	network updates}---i.e., while changing packet-processing rules on network
devices.
In fact, early studies on consistent network updates date back long before the
advent of software-defined networking.
In this section, we provide a historical perspective on the many research
contributions that can be considered as the main precursors of the state of the
art for SDN updates.

We first discuss update problems and techniques in traditional networks
(\S\ref{subsec:history-igp}-\ref{subsec:history-routing}).
In those networks, forwarding rules are computed by routing protocols that run
standardly-defined distributed algorithms, whose output is influenced by both
physical topology (e.g., active links) and routing configurations (e.g., logical
link costs).
Pioneering update works aimed at avoiding transient inconsistencies due to
modified topology or configurations, mainly focusing on the Interior Gateway
Protocols (IGPs) that are commonly used to control forwarding within a single
network.
A first set of contributions tried to modify IGP protocol definitions, mainly to
provide forwarding consistency guarantees upon link or node failures.
Progressively, the research focus has shifted to a more general problem of
finding sequences of IGP configuration changes that modify forwarding while
guaranteeing forwarding consistency, e.g., for service continuity
(\S\ref{subsec:history-igp}).
More recent works have also considered reconfigurations of protocols different
or deployed in addition to IGPs, mostly generalizing previous techniques while
keep focusing on forwarding consistency (\S\ref{subsec:history-routing}).

Subsequently (\S\ref{subsec:history-sdn}), we discuss update problems in
the context of SDNs.
Those networks are based on a clear separation between controller (implementing the
control logic) and dataplane elements (applying controller's decision on
packets).
This separation indisputably provides new flexibility and opens new network design
patterns, for example, enabling security requirements to be implemented by
careful path computation (done by the centralized controller).
This also pushed network update techniques to consider additional consistency
properties like policies and performance.

\begin{figure}[!ht]
   \centering
   \subfloat[Surpassed state]{
      \includegraphics[width=0.3\columnwidth]{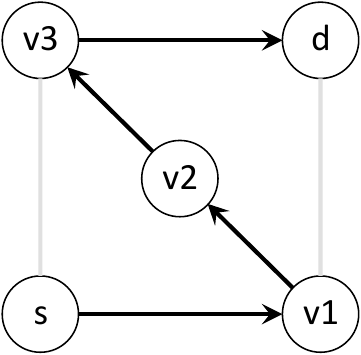}
      \label{subfig:example-history-generic-init}
   }
   \qquad \qquad
   \subfloat[Down state]{
      \includegraphics[width=0.3\columnwidth]{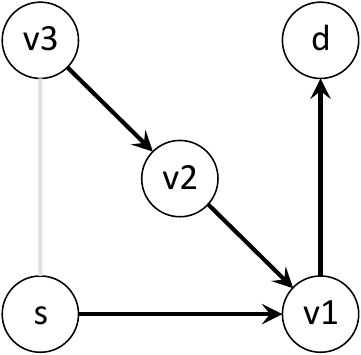}
      \label{subfig:example-history-generic-fin}
   }
   \caption{A network update example, where forwarding paths have to be changed from the Surpassed (Fig.~\ref{subfig:example-history-generic-init}) to the Down (Fig.~\ref{subfig:example-history-generic-fin}) state. Arrows represent paths on which traffic (e.g., from $s$ to $d$) is forwarded, while (gray) undirected links between nodes represent unused links.
   }
   \label{fig:example-history}
\end{figure}

Throughout the section, we rely on the generic example shown in
Fig.~\ref{fig:example-history} for illustration.
The figure shows the intended forwarding changes to be applied for a generic
network update. Observe that possible forwarding loops can occur when we update nodes one by
one, because links $(v1,v2)$ and $(v2,v3)$ are traversed in opposite
directions before and after the update.

\subsection{IGP Reconfigurations}\label{subsec:history-igp}
In traditional (non-SDN) networks, forwarding paths are computed by
distributed routing protocols.
Among them, link-state IGPs are typically used to compute forwarding paths
within a network owned by the same administrative entity.
They are based on computing shortest-paths on a logical view of the network,
that is, a weighted graph which is shared across routers.
Parameters influencing IGP computations, like link weights,
can be set by operators by editing router configurations.

\begin{figure}[h]
   \centering
   \subfloat[Surpassed state]{
      \includegraphics[width=0.3\columnwidth]{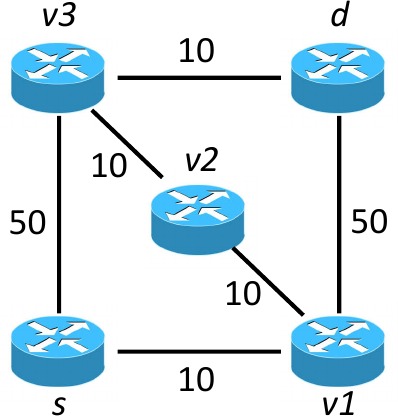}
      \label{subfig:example-history-igp-init}
   }
   \qquad \qquad
   \subfloat[Down state]{
      \includegraphics[width=0.3\columnwidth]{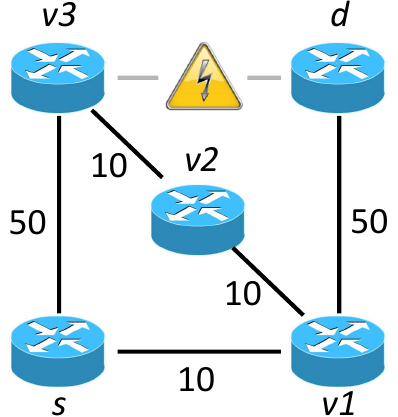}
      \label{subfig:example-history-igp-fin}
   }
   \caption{Possible implementation of pre- and post-update forwarding paths for the update in Fig.~\ref{fig:example-history} in a traditional, IGP-based network. Numbers close to network links represent the corresponding IGP weights.}
   \label{fig:example-history-igp}
\end{figure}

As an illustration, Fig.~\ref{fig:example-history-igp} shows a possible IGP
implementation for the network states shown in Fig.~\ref{fig:example-history}.
In particular, Fig.~\ref{fig:example-history-igp} reports the IGP graph
(consistent with the physical network topology) with explicit mention of the
configured link weights.
Based on those weights, for each destination (e.g., $d$ in this example), all
routers independently compute the shortest paths, and forward the corresponding
packets to the next-hops on those paths. Consequently, the IGP configurations in
Figs.~\ref{subfig:example-history-igp-init}
and~\ref{subfig:example-history-igp-fin} respectively produce the forwarding
paths depicted in Figs.~\ref{subfig:example-history-generic-init}
and~\ref{subfig:example-history-generic-fin}.

When the IGP graph is modified (e.g., because of a link failure, a link-weight
change or a router restart), messages are propagated by the IGP itself
from node to node, so that all nodes rebuild a consistent view of the network:
This process is called \textit{IGP convergence}.
However, IGPs do not provide any guarantee on the timing and ordering in which nodes
receive messages about the new IGP graphs.
This potentially triggers transient forwarding disruptions due to temporary state
inconsistency between a set of routers.
For example, assume that we simply remove link $(v3,d)$ from the IGP graph
shown in Fig.~\ref{subfig:example-history-igp-init}. This will eventually lead
us to the configuration presented in Fig.~\ref{subfig:example-history-igp-fin}.
Before the final state is reached, the notification that $(v3,d)$ is removed
has to be propagated to all routers.
If $v3$ receives such notification before $v2$ (e.g., because closer to the
removed link), then $v3$ would recompute its next-hop based on the new
information, and starts forwarding packets for $d$ to $v2$ (see
Fig.~\ref{subfig:example-history-generic-fin}). Nevertheless, $v2$ keeps
forwarding packets to $v3$ as it considers that $(v3,d)$ is still up. This
creates a loop between $v3$ and $v2$: The loop remains until $v2$ is 
notified about the removed link. A similar loop can occur between $v2$ and
$v1$.

Guaranteeing disruption-free IGP operations has been considered by research and
industry since almost two decades.
We now briefly report on the main proposals in the area, which  share the focus
on support for planned operations.

\smallskip

\subsubsection{Protocol extensions}
Early contributions focused on the modification of IGPs, mainly to
avoid forwarding inconsistencies.
Among them, protocol extensions have been
proposed~\cite{rfc3623,shaikh-graceful_ospf-ton-06,rfc5306} to gracefully
restart a routing process, that is, to avoid forwarding disruptions (e.g.,
blackholes) during the software upgrade of a router.
Other works focused on avoiding forwarding loops during configuration
changes.
For example, Fran{\c{c}}ois and Bonaventure~\cite{ofib-07} propose oFIB, an IGP extension that guarantees the absence of forwarding loops upon manually managed
topological changes, e.g., to propagate information about a link or a router that has
to be shut down for maintenance.
The key intuition behind oFIB is to use explicit synchronization between routers
to constrain the order in which each router changes its forwarding
entries.
Namely, each router is forced not to update its
forwarding entry for a given destination until all its final next-hops switched
to their respective final next-hops for that destination.
Consider again Fig.~\ref{fig:example-history-igp}, assuming that oFIB is deployed.
To prepare the shutdown of link $(v3, d)$, oFIB ensures that $v1$ is the only
router changing its forwarding entry to $d$ at first: this is safe because
$v1$'s final next-hop is directly the destination $d$.
All the other routers (e.g., $v2$) do not update yet until their final next-hops
(e.g., $v1$) use their respective final paths.
In fact, after $v1$ starts forwarding traffic through the $(v1, d)$ link, it
notifies its neighbors about its updated state.
At that point, $v2$ and $s$ can update their respective forwarding entries
for $d$.
The whole network is eventually updated by iterating this process.

oFIB inspired a number of variants, aiming at applying explicit notification
to more generic updates.
Fu \etal\cite{Fu_loop_free_updates_08} generalize\rev{s} the approach by defining a loop-free ordering of IGP-entry updates for arbitrary forwarding
changes.
Shi \etal\cite{Shi_loop_free_congestion_09} also 
extend\rev{s}
 the reconfiguration mechanism to avoid traffic congestion in addition to forwarding incorrectness.
A broader overview of loop avoidance and mitigation techniques mostly inspired
by oFIB is reported 
in~\cite{rfc5715}.

Modifying protocol specifications may seem the most straightforward solution to
deal with reconfigurations in traditional networks, but it actually has practical
limitations.
First, this approach cannot accommodate custom reconfiguration objectives.
For instance, ordered forwarding changes generally work only on a
per-destination basis~\cite{ofib-07}, which can make the reconfiguration process
slow if many destinations are involved -- while one operational objective
could be to exit transient states as quickly as possible.
Second, protocol modifications are targeted to specific reconfiguration cases
(e.g., single-link failures), since it is intrinsically hard to predict the
impact of any possible configuration change on forwarding paths.
Finally, protocol extensions are not easy to implement in practice, because of
the reluctance of vendors to change their proprietary router software, as well
as the additional complexity and potential overhead (e.g., load) induced on
routers.

\smallskip

Limited practicality of protocol modifications quickly motivated new approaches,
based on coordinating operations readily available in deployed routers, at a
per-router level.

\smallskip

\subsubsection{Coarse-grained operation scheduling}
Planned reconfigurations may encompass several coarse-grained operations.
Consider the case where in Fig.~\ref{subfig:example-history-igp-init}, the
weight of link $(s,v1)$ has to be set to $70$ in addition to removing the link
$(v3,d)$.
The link reweighting might be desirable to improve load balancing across the
network, e.g., adapting to a permanently increased volume of traffic from $s$ to
$d$.
Such a reconfiguration effectively consists of two macro operations: removing
$(v3,d)$ and changing the weight of $(s,v3)$.
Even assuming that each coarse-grained operation is atomic, the order in which
distinct operations are performed can have an impact on how much the network
traffic is disrupted during the reconfiguration.
Assume that link $(s,v3)$ can sustain no more than 50\% of the $s-d$ traffic
volume.
Reweighting $(s,v1)$ before removing $(v3,d)$ forces all the $s-d$ traffic on
the path $(s,v3,d)$ which overloads link $(s,v3)$, while removing $(v3,d)$
first would not cause congestion.

Several works propose to use optimization techniques to compute the order of
macro-operations so as to guarantee given consistency properties.
As a first approach, Keralapura \etal\cite{Keralapura_network_upgrade_06} formalized the
problem of finding the optimal order in which nodes can be added to a network,
one by one, so as to minimize an objective function modeling typical costs of
connectivity and traffic disruptions in Internet Service Providers.
T\rev{h}e following contributions encompass additional operations.
In 2009~\cite{raza_gno_infocom_2009} and 2011~\cite{raza_gnsm_ton11}, for
instance, Raza \etal propose a theoretical framework to 
schedule link weight changes in a way that minimizes a generic disruption
function.
This approach enables to formulate our reconfiguration example as a formal
optimization problem, where constraints enforce that the reweighting of links
$(s,v1)$ and $(v3,d)$ (from 10 to 70 and from 10 to $\infty$, respectively)
are both scheduled, and the objective functions aggregates the cost of every
step in the schedule.
The works also describe two algorithms to solve the formalized problems, one
based on dynamic programming and the other on an ant colony optimization
heuristic.

\smallskip

The approaches just described basically spread coarse-grained operations
over time.
This is not sufficient to deal with many update scenarios.
Fig.~\ref{fig:example-history-igp} displays one of such scenarios: since the
reconfiguration includes a single coarse-grained operation (link removal),
previous approaches cannot prevent forwarding loops possibly occurring
when that single operation is performed.

\smallskip

\begin{figure*}[ht]
	\centering
	\subfloat[Initial]{
		\includegraphics[width=0.3\columnwidth]{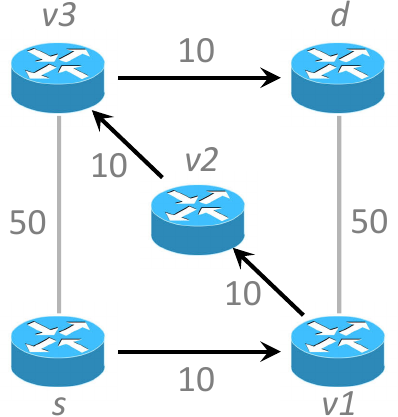}
		\label{subfig:example-history-igp-seq1}
	}\quad
	\subfloat[Step 1]{
		\includegraphics[width=0.3\columnwidth]{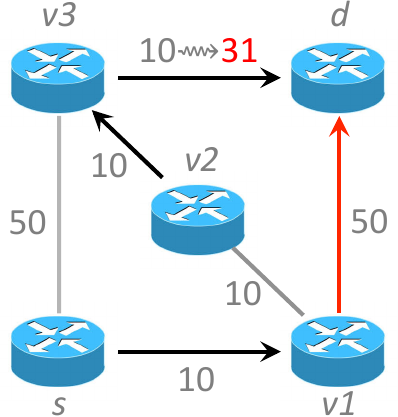}
		\label{subfig:example-history-igp-seq2}
	}\quad	\subfloat[Step 2]{
		\includegraphics[width=0.3\columnwidth]{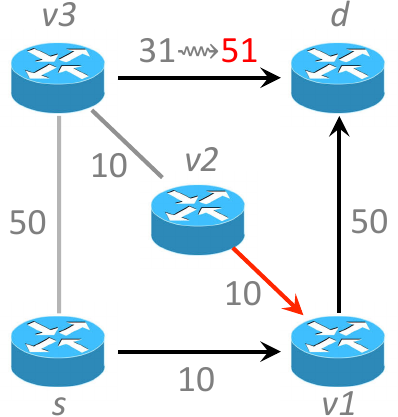}
		\label{subfig:example-history-igp-seq3}
	}\quad
	\subfloat[Final]{
		\includegraphics[width=0.3\columnwidth]{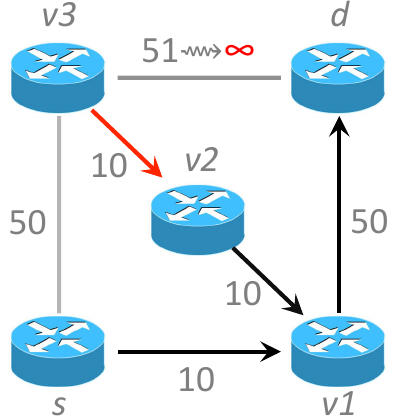}
		\label{subfig:example-history-igp-seq4}
	}
	\caption{Intermediate IGP weights that enable a loop-free reconfiguration for the example in Fig.~\ref{fig:example-history}.}
	\label{fig:example-history-igp-seq}
\end{figure*}

\subsubsection{Progressive link reweighting}
Intermediate link weights can be set to avoid disruptions during a
reconfiguration, even if it includes a single link weight change.
Consider again the example in Fig.~\ref{fig:example-history-igp}, and let the
final weight for link $(v3,d)$ conventionally be $\infty$.
In this case, the forwarding loops potentially triggered by the IGP reconfiguration
can be provably prevented by using two intermediate weights for link $(v3,d)$,
as illustrated in Fig.~\ref{fig:example-history-igp-seq}.
The first of those intermediate weights (see
Fig.~\ref{subfig:example-history-igp-seq2}) is used to force $v1$ and only
$v1$ to change its next-hop, from $v2$ to $d$: Intuitively, this prevents the
loop between $v2$ and $v1$. The second intermediate weight (see
Fig.~\ref{subfig:example-history-igp-seq3}) similarly guarantees that the
loop between $v3$ and $v2$ is avoided, i.e., by forcing $v2$ to use its final
next-hop before $v3$.

Of course, computing intermediate weights that guarantee the absence of
disruptions becomes trickier when multiple destinations are involved.

Such a technique can be straightforwardly applied to real routers. For example,
an operator can progressively change the weight of $(v3,d)$ to $31$ by editing
the configuration of $v3$ and $d$, then check that the all IGP routers have
converged on the paths in Fig.~\ref{subfig:example-history-igp-seq2}, repeat
similar operations to reach the state in
Fig.~\ref{subfig:example-history-igp-seq3}, and finally remove the link safely.
Even better, Fran{\c{c}}ois \etal\cite{fsb-dftron-07} have proved that it is always possible to
compute a sequence of intermediate link weights that provably avoids all
transient loops when a single link has to be reweighted. Obviously, the weight of
multiple links can be changed in a loop-free way, by safely reweighting
links one by one.

Additional research contributions focused on minimizing the number of
intermediate weights that ensure loop-free reconfigurations.
Surprisingly, the problem is \textit{not} computationally hard, despite the fact
that all destinations have potentially to be taken into account when changing
link weights.
Polynomial-time algorithms have been proposed to support planned operations at
the per-link~\cite{clad-link-2013,fsb-dftron-07} (e.g., single-link reweighting)
and at a per-router~\cite{clad-router-icnp2013,clad-router-ton2015} (e.g.,
router shutdown/addition) granularity.

\smallskip

\subsubsection{Ships-in-the-Night (SITN) techniques}
\begin{figure}[h]
   \centering
   \includegraphics[width=0.3\columnwidth]{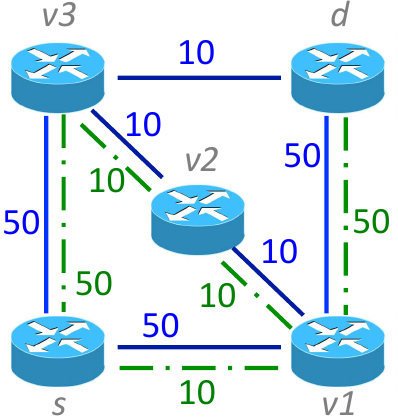}
   \caption{Ships-in-the-Night setup: All routers run two routing processes, one
   with the initial configuration (blue, solid lines) and the other with the
   final configuration (green, dashed and dotted segments).}
   \label{fig:example-history-sitn-setup}
\end{figure}

To improve the update speed in the case of simultaneous link weight changes and
to deal with additional reconfiguration scenarios (from changing routing
parameters like OSPF areas to replacing an IGP with another), both industrial
best practices and research works often rely on a technique commonly called
Ships-in-the-Night~\cite{hv-nmm-10}.
This technique builds upon the capability of traditional routers to run multiple
routing processes at the same time.
Thanks to this capability, both the initial and final configurations can be
installed (as different routing processes) on all nodes at the same time.
When multiple configurations are installed on the same node, only one
of them is preferred and used.
Fig.~\ref{fig:example-history-sitn-setup} shows the setup for a
Ships-in-the-Night reconfiguration for the reconfiguration case in
Fig.~\ref{fig:example-history-igp}.

In SITN, the reconfiguration process then consists in swapping the preference
between the initial and the final configurations on every node, typically one
by one.
Configuration preference can be swapped at a per-destination granularity.
This means that (1) for each destination, every node either forwards packets to
its initial next-hops or its final ones;
(2) at any time during the reconfiguration, distinct nodes can use different
configurations; hence, (3) inconsistencies may arise from the mismatch between
the configurations used by distinct nodes.

\begin{figure*}[thbp]
   \centering
   \subfloat[Initial]{
      \includegraphics[width=0.3\columnwidth]{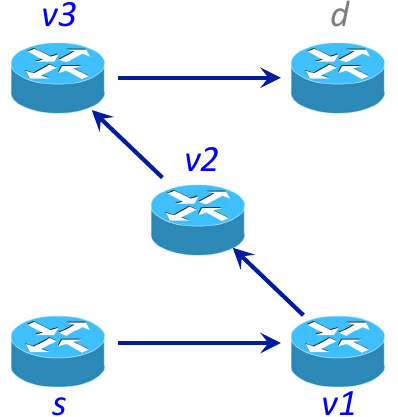}
      \label{subfig:example-history-sitn-seq1}
   }\quad
   \subfloat[Step 1]{
      \includegraphics[width=0.3\columnwidth]{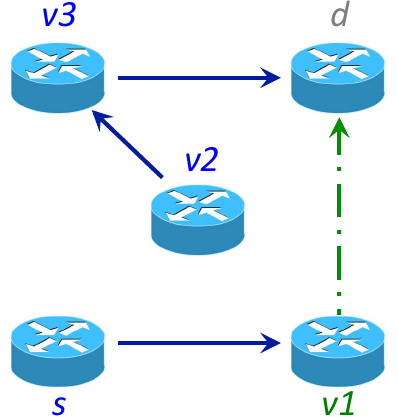}
      \label{subfig:example-history-sitn-seq2}
   }\quad
   \subfloat[Step 2]{
        \includegraphics[width=0.3\columnwidth]{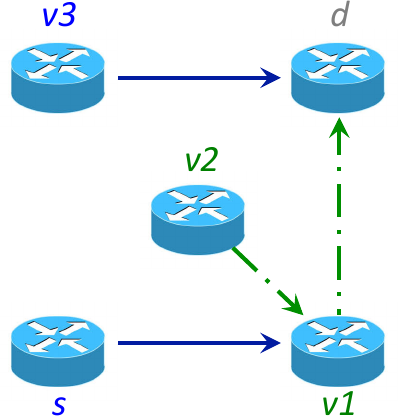}
        \label{subfig:example-history-sitn-seq3}
   }\quad
   \subfloat[Final]{
        \includegraphics[width=0.3\columnwidth]{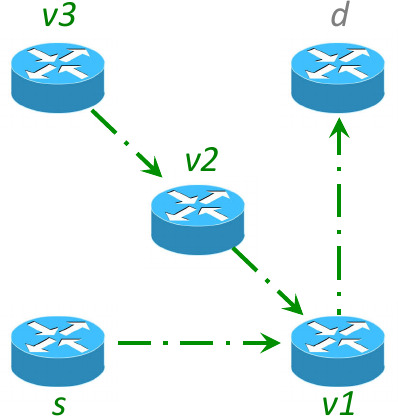}
        \label{subfig:example-history-sitn-seq4}
   }
   \caption{Ships-in-the-Night reconfiguration that mimics the progressive link reweighting shown in Fig.~\ref{fig:example-history-igp-seq}. In each figure, the colors of router names indicate their respective control-plane preferences at the represented reconfiguration step.}
   \label{fig:example-history-sitn-seq}
\end{figure*}

Because of those potential inconsistencies, the Ships-in-the-Night approach
opens a new algorithmic problem, that is, to decide a safe order in which to
swap preferences on a per-router basis.
For example, if the configuration preference is swapped on $v3$ before doing the
same on $v2$ in Fig.~\ref{fig:example-history-sitn-setup}, we end up with a loop
between $v2$ and $v3$.
In contrast, Fig.~\ref{fig:example-history-sitn-seq} shows a SITN-based safe
reconfiguration that mimics the progressive link weight increment depicted in
Fig.~\ref{fig:example-history-igp-seq}.

Naive algorithms for swapping configuration preferences cannot guarantee
disruption-free reconfigurations.
For example, replacing the initial configuration with the final one on all nodes
at once provides no guarantee on the order in which new preferences are applied
by nodes, hence they potentially trigger packet losses and service disruptions
(in addition to massive control-plane message exchanges).
Such an approach will also leave the network in an inconsistent,
disrupted and hard-to-troubleshoot state if any reconfiguration command is lost
or significantly delayed.
Industrial best practices (e.g.,~\cite{hv-nmm-10,p-crpyn-07}) only
provide rules of thumb which do not apply in the general case, and do not
guarantee lossless reconfiguration processes.

Hence, the problem of computing a safe per-router reconfiguration order within
SITN called for new research contributions.
Prominently, \cite{igp-updates,igp-updates-ton} show that no SITN-based update
order guarantees the absence of forwarding loops in some cases, and it is
NP-complete to even assess whether a loop-free order exists.
Those papers also propose two algorithms to compute a loop-free order.
The first algorithm is based on applying traditional optimization algorithms to
solve a Linear Program (LP) that models the input update problem.
Such an LP is derived from enumerating all possible loops, and encoding all the
possibilities to avoid every loop as LP constraints (that must be alternatively
satisfied). For the example in Fig.~\ref{fig:example-history-sitn-setup}, this
algorithm would enumerate the two potential loops (between $v1$ and $v2$, and
between $v2$ and $v3$), and formulate LP constraints forcing the control-plane
preference to be swapped at $v1$ before $v2$ (to break the first loop) and at
$v2$ before $v3$ (to break the second loop).
The second algorithm is a heuristic based on reconfiguring routers according to
the final paths (e.g., $v1$ before $v2$ and $s$ in our example, because $v1$
is closer to $d$), aimed to avoid the scalability problems of the LP-based
approach.
Finally, \cite{igp-updates,igp-updates-ton} describe a comprehensive system to
carry out loop-free SITN-based reconfigurations.
The system computes the operational sequence to perform an input
reconfiguration, and directly interacts with the routers to modify their
configurations, and check when every operation in the sequence is completed.
The system was envisioned to work semi-automatically, waiting for an explicit
confirmation from the operator before performing the next operation in the
computed sequence.

\subsection{Generalized Routing Reconfigurations in Traditional\\ $~~~~~$Networks}\label{subsec:history-routing}

Research contributions have been devoted to reconfigurations in more realistic
settings, including other protocols in addition to an IGP.

\smallskip

\subsubsection{Enterprise networks, with several routing domains}
As a first example, the Ships-in-the-Night framework has been used to carry out
IGP reconfigurations in enterprise networks. Those networks typically use
\textit{route redistribution}~\cite{route-redistribution-2008}, a mechanism
enabling the propagation of information from one routing domain (e.g., running
an IGP) to another (e.g., running another IGP). Route redistribution may be
responsible for both routing (inability to converge to a stable state) and
forwarding (e.g., loop) anomalies~\cite{route-redistribution-2008}. SITN-based
update procedures have been proposed in~\cite{route-redistribution-2014} to
avoid transient anomalies while (i)~ reconfiguring a specific routing domain,
and/or (ii)~ arbitrarily changing the size and shape of routing domains.

\smallskip

\subsubsection{Internet Service Providers (ISPs), with BGP and MPLS}
In ISP networks, the Border Gateway Protocol (BGP) and often the Multi-Protocol
Label Switching (MPLS) protocol are pervasively used to manage transit
traffic, for which both the source and the destination is external to the
network. Vanbever \etal\cite{igpbgp-2013} showed that even techniques guaranteeing safe IGP reconfigurations can cause transient
forwarding loops in those settings, because of the interaction between IGP and
BGP.
They also proved conditions to avoid those BGP-induced loops during IGP
reconfigurations, by leveraging the presence of MPLS or carefully configuring
BGP (according to some guidelines).

In parallel, a distinct set of techniques aimed at supporting BGP
reconfigurations.
Fran{\c{c}}ois \etal\cite{BGP-session-maintenance-07} propose a solution to avoid churn and loss of
connectivity due to planned BGP session shutdown: This solution is based on
modifying BGP to distribute alternate routes and move traffic on them
\textit{before} the target BGP session is actually removed.
Wang \etal\cite{wang-vroom-08} present an approach, based on extending virtual machine
migration techniques, to quickly transfer virtual routers from one physical device to
another.
Keller \etal\cite{keller_bgp_grafting_10} address the more general problem of fastly
migrating parts of the BGP configuration (e.g., transferring a BGP session from
one router to another), with a technique that takes care of moving the BGP state
to the new route and reduce the impact of the migration on both BGP peers and
other routers.
Vissicchio \etal\cite{bgp-reconfigs-ton13} describe a framework that enables radical
re-organizations of BGP sessions (e.g., changing several of them, or
transforming a full-mesh into a route reflector topology) while guaranteeing the
absence of forwarding and routing anomalies: The framework is based on
implementing Ships-in-the-Night in BGP (with a minimal extension to existing
routers), and tagging packets so that routers can uniformly apply a single BGP
configuration to every packet.

Finally, Internet-level problems, like maintaining global connectivity upon
failures, have also been explored (see, e.g.,~\cite{RBGP-07}).

\smallskip

\subsubsection{Protocol-independent reconfiguration frameworks}
By design, all the above approaches are dependent on the considered (set of)
protocols and even on their implementation.

Protocol-independent reconfiguration techniques have been studied 
as well in the literature. Mainly,
\cite{alimi_shadow_config_2008}
generalizes SITN by
proposing a new design for the internal router architecture. This re-design
would allow routers not only to run multiple configurations simultaneously, and
to select the configuration to apply for every packet on the basis of a specific
bit in the packet header. The work also describes a commit protocol to support
the switch between configurations without creating forwarding loops.
General mechanisms for consensus routing have also been explored
in~\cite{ref21}.

\subsection{\rev{Updates of Software-Defined Networks}}\label{subsec:history-sdn}

Recently, software-defined networking has grown in popularity, thanks to its
promises to spur abstractions, mitigate compelling management problems and avoid
network ossification~\cite{openflow}. 
SDN is currently used or discussed in 
a wide range of contexts~\cite{DBLP:journals/pieee/KreutzRVRAU15}, e.g., to improve network virtualization in datacenters, generalize traffic engineering in the wide-area network, or enable slicing in emerging 5G applications, to just name a few.

In pure
SDN networks, rather than having
devices (switches and routers) run their own distributed control logic,
the controller computes (according to operators' input) and
installs (on the controlled devices) the rules to be applied to packets
traversing the network: No message exchange or distributed computation are
needed anymore on network devices.
This is very different from the existing decentralized control planes typically used in traditional networks (as well as in many Ad Hoc and P2P networks)
and allows, e.g., to overcome the notoriously slow reaction
to changes (e.g., link failures)
and rerouting of flows in those networks: one of the key reasons behind Google's move to SDN~\cite{google-move}.

\begin{figure}[ht]
   \centering
	\includegraphics[width=0.60\columnwidth]{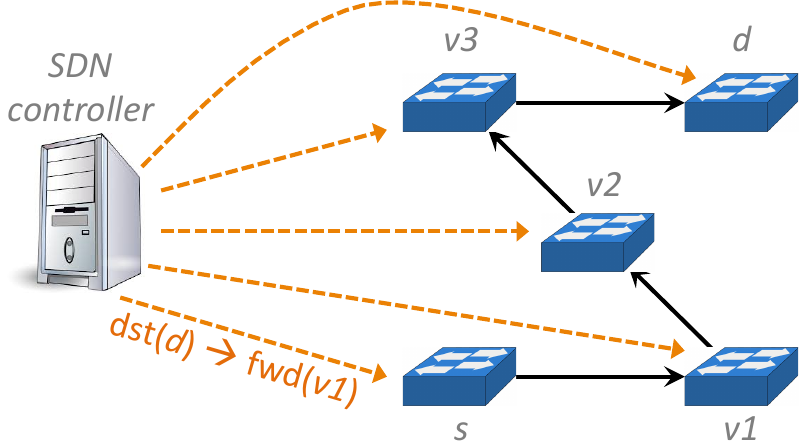}
   \caption{Implementation of the surpassed state in
   Fig.~\ref{fig:example-history} in an SDN network.}
   \label{fig:example-history-sdn-init}
\end{figure}

Fig.~\ref{fig:example-history-sdn-init} depicts an example of an SDN network,
configured to implement the initial state of our update example (see
Fig.~\ref{fig:example-history}).
Beyond the main architectural components, the figure also 
illustrates a classic
interaction between them.
Indeed, the dashed lines indicate that the SDN controller instructs the
programmable network devices, typically switches~\cite{openflow}), on how to
process (e.g., forward) the traversing packets.
An example command sent by the controller to switch $s$ is reported next to
the dashed line connecting the two: This command instructs $s$ to use $v1$
as next-hop for any packet destined to $d$.

The SDN architecture is expected to make network updates more frequent and
more critical than in traditional networks.
On the one hand, controllers are often intended to support several different
requirements, including performance (like optimal choice of per-flow paths),
security (like firewall and proxy traversal) and packet-processing (e.g.,
through the optimized deployment of virtualized network functions) ones.
On the other hand, devices cannot provide any reaction (e.g., to topological
changes) like in traditional networks.
In turn, this comes at the risk of triggering inconsistencies, e.g., creating
traffic blackholes during an update, that are provably impossible to trigger by
reconfiguring current routing protocols~\cite{coexistence-controlplanes-15}. As
a consequence, the controller has to carry out a network update for every event
(from failures to traffic surges and requirement modification) that can impact
the forwarding rules installed on the switches; additionally, it should perform
such updates while typically supporting more critical consistency guarantees
(e.g., security-related ones) and performance objectives (e.g., for prompt
reaction to failures) than in traditional networks.

An extended corpus of SDN update techniques have already been proposed in the
literature, following up on the large interest raised by SDN
in the last few years.
This research effort nicely complements approaches to
specify~\cite{sdn-languages-survey-13}, compile~\cite{ethane,pyretic}, and
check the implementation of~\cite{realTimeCheck,hsa} network
requirements that operators may want to implement in their (SDN) networks.

\begin{figure*}[thbp]
   \centering
   \subfloat[Step 1]{
      \includegraphics[width=0.6\columnwidth]{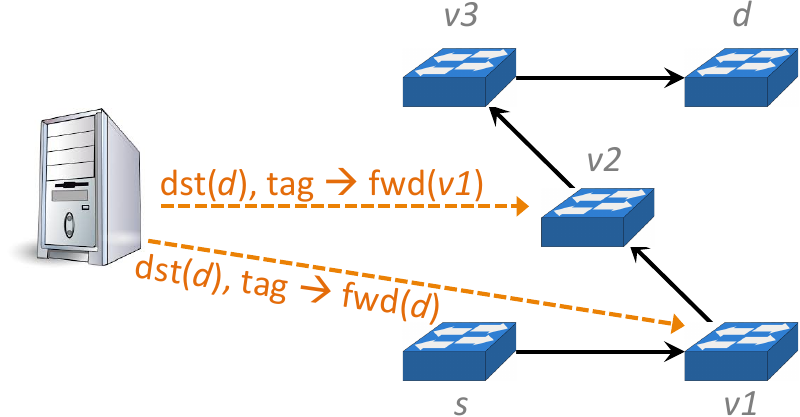}
      \label{subfig:example-history-2phase-seq1}
   }
   \hfill 
   \subfloat[Step 2]{
			\includegraphics[width=0.6\columnwidth]{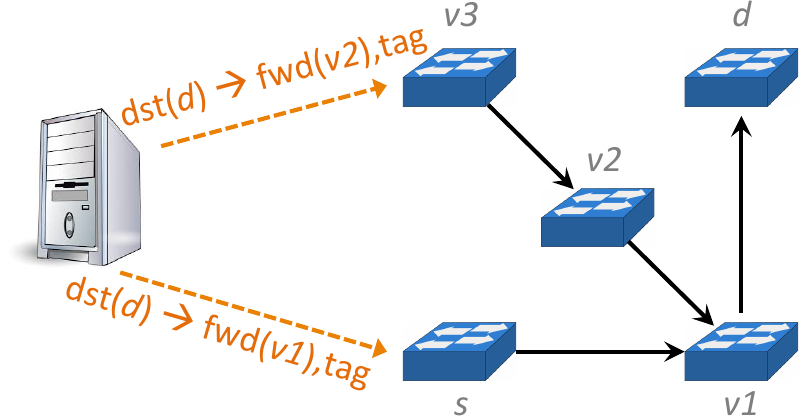}
      \label{subfig:example-history-2phase-seq2}
   }
   \hfill
   \subfloat[Step 3]{
				\includegraphics[width=0.6\columnwidth]{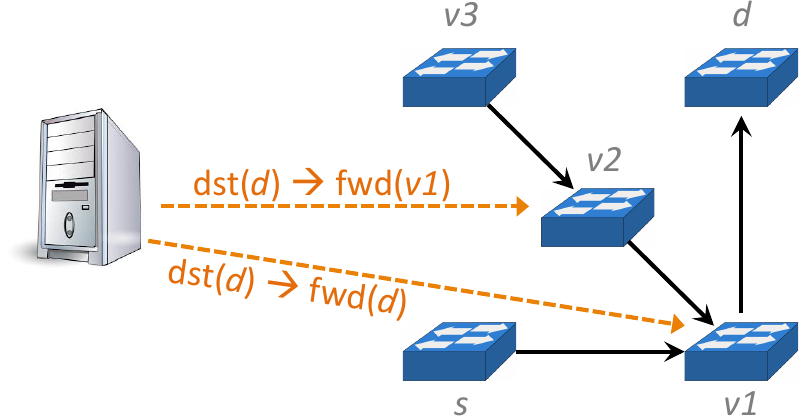}
        \label{subfig:example-history-2phase-seq3}
   }
   \caption{Application of the 2-phase commit technique for carrying out our update example (see Fig.~\ref{fig:example-history-igp-seq}). A final (optional) step consists in cleaning the configuration by removing packet tags, i.e., reverting tagging at $v3$ and $s$ as enforced by Step 2.}
   \label{fig:example-history-2phase-seq}
\end{figure*}

The first cornerstone of SDN updates is represented by
the work by Reitblatt \etal in 2011~\cite{jaq4} and 2012~\cite{abstractions}.
This work provides a first analysis of the additional (e.g., security)
requirements to be considered for SDN updates, extending the scope of
consistency properties from forwarding to policy ones. In particular, it
focuses on per-packet consistency property, imposing that packets have to be
forwarded either on their initial or on their final paths (never 
a combination of the two), throughout an update.

The technical proposal is centered around the 2-phase commit technique,
which relies on tagging packets at the ingress so that either all
initial rules or all final ones can be consistently applied network-wide.
Initially, all packets are tagged with a given ``old label'' (e.g., no tag) and
rules matching the old label are pre-installed on the switches. In a first step,
the controller then instructs the internal switches to apply the final
forwarding rules to packets carrying a ``new label'' -- even if no packet
carries such label at this step.
After the internal switches have confirmed the successful installation of
these new rules, the controller then changes the tagging policy at the ingress
switches, requiring them to tag packets with the ``new label''.  As a result,
packets are immediately forwarded along the new paths. Finally, the internal
switches are updated (to remove the old rules), and an optional cleaning step
can be applied to remove all tags from packets.
Fig.~\ref{fig:example-history-2phase-seq} shows the operational sequence
produced by the 2-phase commit technique for the update case in
Fig.~\ref{fig:example-history-igp-seq}.

Several works have been inspired by the 2-Phase technique presented
in~\cite{abstractions}.
One first line of research focuses on providing additional guarantees, e.g., congestion-freedom
(from~\cite{swan} to~\cite{SebastianBrandt2016,roger-infocom,Forster2016Power,dionysus,zupdate,DBLP:conf/icc/LuoYLZ15,icnp-jiaqi}).
In a second line of research, several algorithms~\cite{igp-updates-ton,dsn16,Forster2016Consistent,sigmetrics16,8214220} to compute a set of ordered
rule replacements have been proposed to deal with specific SDN update cases
(e.g., where only forwarding consistency is needed), avoiding the need for additional rules and hence enabling more resource-efficient approaches (e.g., TCAM memory slots are expensive and precious).

In the following sections, we detail most of those contributions and the
insights on different update problems that globally emerge from them.

For a tutorial description of a few major works in the area, mainly~\cite{dionysus,roger},
 we refer to the recent article of Li \textit{et al}.~\cite{DBLP:journals/fcsc/LiWZX17}.

\section{\rev{Taxonomy of Update Techniques}}\label{sec:taxo}

We now present a general formulation of network update
problem (\S\ref{subsec:taxo-problem}), which abstracts from assumptions and
settings considered in the literature.
This formulation enables us to classify research contributions on the basis of
the proposed techniques (e.g., simultaneous usage of multiple configurations on
nodes or not) and algorithms, independently of their application to traditional
and SDN networks (\S\ref{subsec:taxo-taxo}).

\subsection{\rev{Generalized Network Update Problem (GNUP)}}\label{subsec:taxo-problem}

In order to compare and contrast research contributions, we first provide a
generalized statement for network update problems.
We use again Fig.~\ref{fig:example-history} for illustration.

\smallskip

\subsubsection{Basic Problem}
Generally speaking, a network update problem consists in computing a sequence of
operations that changes the packet-processing rules installed on network
devices.
Consider any communication network: It is composed by a given set of
inter-connected devices, that are able to process data packets (e.g., forwarding
them to a next-hop) according to rules installed on them.
We refer to the set of rules installed on all devices at a given time as network
state at that time.
Given an initial and final state, a network update consists in passing from the
initial state to the final one by applying operations (i.e., adding, removing or
changing rules) on different devices.
In Fig.~\ref{fig:example-history}, the initial state forces packets from source $s$
to destination $d$ along the path $(s,v1,v2,v3,d)$; the final state
forwards the same packets over $(s,v1,d)$, and packets from $v3$ to
$d$ on $(v3,v2,v1,d)$. The network update problem consists in replacing the
initial rules with the final ones, so that the paths for $d$ are updated from
$(s,v1,v2,v3,d)$ to $(s,v1,d)$ and $(v3,v2,v1,d)$.

\smallskip

\subsubsection{Operations}
To perform a network update, a sequence of operations has to be computed.
By operation, we mean a (direct or indirect) modification of packet-processing
rules installed on one or more devices.
As an example, an intuitive and largely-supported operation on network
devices is rule replacement, which consists in instructing a device
(e.g., $v3$) to replace an initial rule (e.g., forward the $s-d$ packet flow to
$v2$) with the corresponding final one (e.g., forward the $s-d$ flow to $d$).
Operations can be coarse-grained and indirect, as IGP link reweighting or
configuration swapping in legacy networks that imply multiple rule replacements
at distinct devices (see~\S\ref{sec:history}).

\smallskip

\subsubsection{Consistency}
The difficulty in solving network update problems is that some form of
consistency must be guaranteedly preserved \emph{during} the update, for practical
purposes (e.g., avoiding service disruptions and packet losses).
Preserving consistency properties, in turn, depends on the order in which
operations appear in the computed sequence and are executed by network devices.
For example, if $v3$ replaces its initial rule with its final one before $v2$
in Fig.~\ref{fig:example-history}, then the operational sequence triggers a
forwarding loop between $v2$ and $v3$ that interrupts the connectivity from
$s$ to $d$.
In \S\ref{subsec:taxo-taxo}, we provide an overview of consistency properties
considered in the literature.

The practical need for guaranteeing consistency has two main consequences
(as shown in \S\ref{sec:history}).
First, it forces network updates to be performed incrementally, i.e.,
appropriately scheduling operations over time so that the installed
intermediate states are provably disruption-free.
Second, it requires a careful computation of operational sequences, implementing
specific reasoning in the problem-solving algorithms (e.g., to avoid replacing
$v3$'s rule before $v2$'s one in the previous example).

\smallskip

\subsubsection{Performance}
Another algorithmic challenge consists in optimizing network-update performance.
As an example, minimizing the time to complete an update is commonly considered
among those optimization objectives.
Indeed, carrying out an update incrementally requires to install intermediate
configurations, and in many cases it is practically desirable to minimize the
time spent in such intermediate states.
We provide a broader overview of performance goals considered by previous works in
\S\ref{subsec:taxo-taxo}.

\smallskip

\subsubsection{Final Operational Sequences}
Generally, the solution for an update problem can be represented as a sequence
of \textit{steps} or \textit{rounds}, that both (i) guarantees consistency
properties and (ii) optimizes update performance.
Each step is a set of operations that can be started at the same time.
Note that this does not mean that operations in the same step are assumed to be
executed simultaneously on the respective devices;
Rather, all operations in the same step can be started in parallel because the
target consistency properties are guaranteed independently of the relative
order in which those operations are executed.
Examples of operational sequences, computed by different techniques, are
reported in \S\ref{sec:history} (see Figs.~\ref{fig:example-history-igp-seq}
and~\ref{fig:example-history-sitn-seq}).

\subsection{\rev{Classifying Update Techniques According to the Addressed GNUP Instance}}\label{subsec:taxo-taxo}

\begin{figure}[t]
\scalebox{0.9}{
	\centering
	\begin{tikzpicture}[auto]
	\node (top) at (0.5,0) [draw, rounded corners=0.1cm,fill={rgb:black,0.2;white,2}]{\footnotesize \text{Taxonomy of Consistent Network Update Problems}};
	
	\node (concon) at (-2.8,-1.25) [draw, rounded corners=0.1cm,fill={rgb:black,0.2;white,2}]{\scriptsize Connectivity Consistency~\rev{\cite{roger}}};
	\node (polcon) at (0.5,-1.25) [draw, rounded corners=0.1cm,fill={rgb:black,0.2;white,2}]{\scriptsize Policy Consistency~\rev{\cite{jaq4}}};
	\node (capcon) at (3.7,-1.25) [draw, rounded corners=0.1cm,fill={rgb:black,0.2;white,2}]{\scriptsize Capacity Consistency~\rev{\cite{swan}}};
	
	\node (loop) at (-3.4,-2.7) [draw, rotate=90, rounded corners=0.1cm,fill={rgb:black,0.2;white,2}]{\scriptsize Loop Freedom~\rev{\cite{8214220}}};
	\node (black) at (-2.4,-2.95) [draw, rotate=90, rounded corners=0.1cm,fill={rgb:black,0.2;white,2}]{\scriptsize Blackhole Freedom~\rev{\cite{Forster2016Consistent}}};
	
	\node (perpa) at (-0.0,-3.15) [draw, rotate=90, rounded corners=0.1cm,fill={rgb:black,0.2;white,2}]{\scriptsize Per-Packet Consistency~\rev{\cite{abstractions}}};
	
	\node (perpa) at (1.0,-3.15) [draw, rotate=90, rounded corners=0.1cm,fill={rgb:black,0.2;white,2}]{\scriptsize Waypoint Enforcement~\rev{\cite{hotnets14update}}};
	
	\node (congaw) at (3.2,-2.92) [draw, rotate=90, rounded corners=0.1cm,fill={rgb:black,0.2;white,2}]{\scriptsize Congestion-Aware~\rev{\cite{roger-infocom}}};
	
	\node (lataw) at (4.2,-2.75) [draw, rotate=90, rounded corners=0.1cm,fill={rgb:black,0.2;white,2}]{\scriptsize Latency-Aware~\rev{\cite{DBLP:conf/icdcs/ZhengLTFSCW18}}};

	\draw [thick,decorate,decoration={brace,amplitude=5pt},rotate=180] (-1.8,5) to (4.3,5);
	
	\draw [thick,decorate,decoration={brace,amplitude=5pt},rotate=180] (-5.3,5) to (-2.3,5);
	
	\node (cfo) at (3.8,-6.00) [draw, rounded corners=0.1cm,fill={rgb:black,0.2;white,2}]{\scriptsize \text{Cross-Flow Objectives}~\rev{\cite{dionysus}}};
	
	\node(lbo) at (-2.98,-6.50) [draw, rounded corners=0.1cm,fill={rgb:black,0.2;white,2}]{\scriptsize \text{Link-Based Objectives}~\rev{\cite{saeed-arxiv}}};
	
	\node (rbo) at (0.4,-6.50) [draw, rounded corners=0.1cm,fill={rgb:black,0.2;white,2}]{\scriptsize \text{Round-Based Objectives}~\rev{\cite{DBLP:conf/nca/FoersterJSW16}}};
	
	\node (aug) at (3.1,-7.00) [draw, rounded corners=0.1cm,fill={rgb:black,0.2;white,2}]{\tiny \text{Augmentation}~\rev{\cite{icnp-jiaqi}}};
	\node (touch) at (4.7,-7.00) [draw, rounded corners=0.1cm,fill={rgb:black,0.2;white,2}]{\tiny \text{Touches}~\rev{\cite{dsn16-new}}};

	\draw [markovedge, thick] (top) to (concon.north);
	\draw [markovedge, thick] (top) to (polcon.north);
	\draw [markovedge, thick] (top) to (capcon.north);
	
	\draw [markovedge, thick] (cfo) to (aug.north);
	\draw [markovedge, thick] (cfo) to (touch.north);
	
	\draw [markovedge, thick] (3.8,-5.2) to (cfo.north);
	\draw [markovedge, thick] (3.8,-5.2) to (lbo.north);
	\draw [markovedge, thick] (3.8,-5.2) to (rbo.north);
	
	\draw [markovedge, thick] (-1.25,-5.2) to (rbo.north);
	\draw [markovedge, thick] (-1.25,-5.2) to (lbo.north);

	\end{tikzpicture}
	}
	\caption{Types of consistent network update problems, defined independently of the network setting (rule granularity and supported operations).}	\label{fig:taxonomy}
\end{figure}
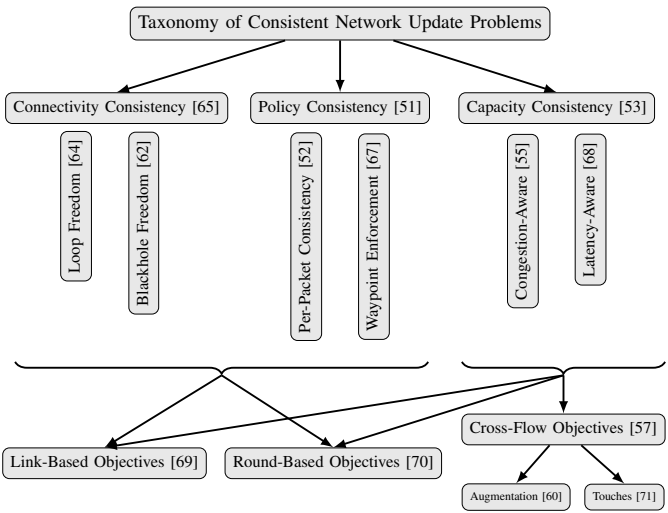

In this section, we provide an overview of the problem space and classify
existing models and techniques.
Previous contributions have indeed considered several variants of the
generalized network update problem as we formulated in
\S\ref{subsec:taxo-problem}.
Those variants basically differ in terms of the update problem on which they
focus. Update problems in turn define both (1) the network setting, including
admitted rule granularity and operations, (2) consistency properties, and (3)
performance goals. An overview of the main update problems considered 
previously
 is
depicted
 in Fig.~\ref{fig:taxonomy}, where we skipped
 the 
orthogonal network
setting dimensions, rule granularity and supported operations, for clarity.

\smallskip
\subsubsection{Rule Granularity}
Network update techniques assume that the underlying devices support rules that
implement one of the two alternative routing models:
\emph{destination-based} and \emph{per-flow} routing.
\paragraph{Destination-based Routing}
In destination-based routing, routers forward packets based on
the destination only. An example for destination-based routing is 
IP routing, where routers
forward packets based on the longest common IP destination prefix. 
In particular, destination-based routing describes confluent paths:
once two flows from different sources destined toward the same destination
intersect at a certain node, the remainder (suffix) of their paths will
be the same.
In destination-based routing, routers store at most one forwarding
rule per specific destination.

\paragraph{Per-flow Routing}
In contrast, according to \emph{per-flow} routing, routes are 
not necessarily confluent: the forwarding rules at the routers
are defined per-flow, i.e., they may depend not only on the destination
but for example also on the source. In traditional networks,
flows and per-flow routing could
for example be implemented using MPLS: packets belonging to the same
equivalence class respectively packets with the same MPLS tag are forwarded
along the same path.

\smallskip
\subsubsection{Operations}
Techniques to carry out network updates can be classified in broad categories,
depending on the operations that they consider.
\paragraph{Rule replacements}
A first class of update techniques is based on computing an order in which
initial rules are replaced by the corresponding final ones on the devices.
Depending on the target setting (e.g., legacy networks or SDNs), such
replacement can be admitted at different granularity, i.e., on a per-rule and
per-device basis (as in OpenFlow networks) or on a per-group of rules and
devices (link reweighting in IGP).

\paragraph{Rule additions}
A second class of network update algorithms is based on adding rules to guarantee
consistency during the update.
The following two main variants of this approach have been explored so far.

\hspace{1.1cm} \textit{1) 2-Phase commit:}
In this case, both the initial and the final rules are installed on all devices
in the central steps of the updates. Packets are tagged at the border of the
network to enforce that the internal devices either (i)~ all use the initial
rules, or (ii)~ all use the final rules. See Fig.~\ref{fig:example-history-2phase-seq}
for an example.

\hspace{1.1cm} \textit{2) Helper rules:}
Some techniques introduce additional rules, which do not belong neither to the
old state nor to the new one, in some intermediate update step. These
rules allow to divert the traffic temporarily to other parts of the network, and
are called \emph{helper rules}.

\paragraph{Mixed}Recently, some update techniques combine rule replacements
and additions, in order to reduce the update overhead (especially in terms of
device-memory consumption) while keeping the flexibility provided by adding
rules.

\smallskip

\subsubsection{Consistency properties}
Update techniques typically target to preserve one (or more) of the following
consistency properties.

\paragraph{Connectivity consistency}
The most basic form of consistency regards the capability of the network to
keep delivering packets to their respective destinations, throughout the
update process.
This boils down to guaranteeing two correctness properties: absence of
blackholes (i.e., paths including routers that cannot forward the packets
further) and absence of forwarding loops (i.e., packets bouncing back and forth
on a limited set of routers, without reaching their destinations).

\paragraph{Policy consistency}
Paths used to forward packets may be selected according to specific forwarding
policies, for example, security ones imposing that given traffic flows must
traverse specific waypoints (firewalls, proxies, etc.).
In many cases, those policies have to be preserved during the update.
Generally speaking, policy consistency properties impose constraints on which
paths can be installed during the update.
For example, an already-mentioned policy consistency property (see
\S\ref{sec:history}) is \textit{per-packet consistency}, requiring that packets
are always forwarded along either the pre-update or the post-update paths, but
never a combination of the two.

\paragraph{\Congestion~consistency}
A third class of consistency properties takes into account the actual availability
and limits of network resources. For instance, many techniques
account for traffic volumes and corresponding constraints raised by the limited
capacity of network links: Those techniques aim at respecting link-capacity
constraints in each update step, e.g., to avoid \textit{transient congestion}
during updates.

\smallskip
\rev{
\textit{Note:} As mentioned in the introduction, most work on SDN network updates argues for
the need to preserve the considered consistency properties at each and every
moment during an update.
In the following sections, we therefore assume a strong consistency model.
We survey approaches relaxing this consistency model beginning in \S\ref{sec:discussion}.
}

\smallskip

\subsubsection{Performance goals}
We can distinguish between three broad classes of 
performance goals.

\paragraph{Link-based}
A first class of consistent 
network update protocols
aims to make new links available as soon as possible,
i.e., to maximize the number of switch rules
which can be updated simultaneously without violating
consistency. 

\paragraph{Round-based}
A second class of consistent network update protocols aims to minimize the total makespan, by computing a schedule of rounds or steps, each consisting of switch rules that are safe to update simultaneously.

\paragraph{Cross-Flow Objectives}
A third class of consistent network
update protocols targets objectives
arising in the presence of multiple flows.

\hspace{1.1cm} \textit{1) Augmentation:}
Minimize
the extent to which link capacities
are oversubscribed during the update
(or make the update entirely congestion-free).

\hspace{1.1cm} \textit{2) Touches:}
 Minimize the number
of interactions with the switch, i.e., the sent messages.

\smallskip 
\rev{
\textit{Note:}} Link-based and round-based objectives are
usually considered for node-ordering algorithms
and
for weak-consistency models.
Congestion-based objectives
are naturally considered for 
capacitated consistency models. 

\rev{
\subsection{Summary and Insights}
We formulated a generalized version of consistent network update problems studied by prior work.
This generalization enables us to create a taxonomy of network update techniques, where previous contributions are classified along four dimensions: assumed rule granularity, operations allowed on the switches, consistency properties to be preserved and optimization goals.
We structure this survey according to the dimension of the consistency properties, because of its importance in the definition of the addressed update problem: the following sections reflect this choice. 

}

\section{\rev{Update Techniques to Guarantee\\Connectivity Consistency}}\label{sec:forwarding}

In this section, we focus on update problems where 
the main consistency property to be guaranteed
concerns the delivery of packets to their
respective destinations.
Packet delivery can be disrupted during an update by forwarding loops or
blackholes transiently present in intermediate states.
We separately discuss previous results on how to guarantee loop-free and
blackhole-free network updates.
We start from the problem of avoiding forwarding loops during updates, because
they are historically the first update problems considered -- by works on
traditional networks (see \S\ref{sec:history}).
This is also motivated by the fact that blackholes cannot be created by reconfiguring
current routing protocols, as proved in~\cite{coexistence-controlplanes-15}.
We then shift our focus on avoiding blackholes during arbitrary (e.g., SDN)
updates.

\subsection{\rev{Guaranteeing Loop-Freedom}}

Loop-freedom is a most basic consistency property and has hence been explored intensively already in the network update literature.
So far in this work, we presented the notion of loop-freedom in the following framework: 1) routing based on the destination, and 2) avoiding all transient loops.
Current research extends the loop-freedom model in both dimensions, introduced next, beginning with the routing model.

\subsubsection{Definitions}

We distinguish between flow-based and destination-based routing:
in the former, we can focus on a single (and arbitrary) path
from $s$ to $d$: forwarding rules stored in the switches
depend on both $s$ and $d$, and flows can be considered
independently. In the latter, switches store a single forwarding
rule for a given destination: once the paths of two different sources
destined to the same destination
intersect, they will be forwarded along the same nodes
in the rest of their route: the routes are confluent.

\begin{figure}[t]
   \centering
   \subfloat[Initial]{
      \includegraphics[width=0.79\columnwidth]{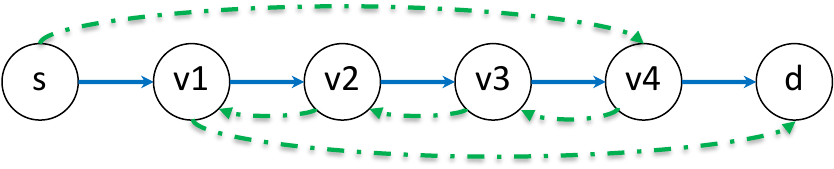}
      \label{subfig:strong1}
   }
   \qquad \qquad
   \subfloat[After updating $s$]{
      \includegraphics[width=0.79\columnwidth]{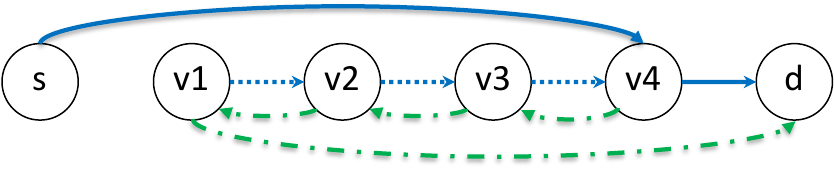}
      \label{subfig:strong2}
   }
   \caption{Example to illustrate the differences between strong and relaxed loop-freedom. The current rules are drawn in solid blue, the new rules are drawn in dash-dotted green. In the left initial state, the task is to update all nodes to use the new forwarding rules. The right figure shows the network state after updating the node s. Note that the current rules for $v1,v2,v3$ are no longer on a path from $s$ to $t$, and are hence drawn dotted. Thus, under RLF, $v1, v2, v3$ can be updated next, as possible loops are not connected to the source. Using SLF, $v3$ must be updated \emph{after} $v2$, which in turn must be updated \emph{after} $v1$. Hence, by updating $v4$ last, the RLF schedule length is just three, even when the construction is extended from 4 nodes
	to up to $vx$ nodes. On the other hand, SLF requires at least $x$ rounds. Hence, RLF can yield a speedup linear in the number of nodes.}
   \label{fig:strongrelax}
\end{figure}

Moreover, one can distinguish between two different definitions
for loop-free network updates: \emph{Strong Loop-Freedom (SLF)}
and \emph{Relaxed Loop-Freedom (RLF)}~\cite{8214220}.
SLF requires that at any point in time, the forwarding rules stored
at the switches should be loop-free. RLF only requires
that forwarding rules stored by switches \emph{along the path from a source $s$
to a destination $d$} are loop-free: only a small number of
``old packets'' may temporarily be forwarded along loops.
RLF can significantly speed up the consistent migration process, as illustrated in Fig.~\ref{fig:strongrelax}.

\subsubsection{Algorithms and Complexity}

Two performance objectives are investigated in the literature, node-based und round-based, to be discussed in turn. Node-based objectives were studied first
in the literature: the goal is to update as many nodes/links at a time as possible. Round-based objectives can be seen as more intuitive, aiming at minimizing the number of (controller-switch interaction) rounds. Node-based approaches are also used in round-based contexts, with the following intuition: by updating as many nodes as possible, the total makespan is hopefully minimized -- coining the notion of \emph{greedy approaches}. However, it has been shown that node- and round-based objectives can conflict and lead to vastly different schedules.

\smallskip
\paragraph{Node-based objective (``greedy approach'')}
Mahajan and Wattenhofer~\cite{roger} initiated the study of destination-based (strong) loop-free network updates.
In particular, the authors show that by scheduling updates based on combinatorial dependencies, consistent update schedules can be derived which do not require any packet tagging, and which allow some updated links to become available earlier. 
The authors also present a first algorithm that quickly updates routes in a transiently loop-free manner: based on the current state, the controller greedily attempts to update as many nodes as possible. 
E.g., in Fig.~\ref{subfig:strong1}, their algorithm picks the nodes $s$ and $v1$, as $v2$-$4$ have unresolved dependencies.
The study of this model has been refined in~\cite{Forster2016Consistent,Forster2016Power}, where the authors also establish hardness results, see Table~\ref{loop-table}.
A related variant using so-called proof labeling schemes~\cite{DBLP:journals/dc/KormanKP10} was proposed in~\cite{DBLP:journals/tcs/FoersterLSW18}.

\begin{table*}[htbp]
\caption{Overview of results for loop-freedom.}
\label{loop-table}
\centering
\scriptsize
\begin{tabular}{|>{\centering}p{1.2cm}|>{\centering}p{4cm}|>{\centering}p{4cm}|>{\centering}p{7cm}|}
\hline
 Model & NP-hard & Polynomial time & Remarks \tabularnewline 
\hline  
\hline
\# Rounds, strong LF & Is there a 3-round loop-free update schedule?~\cite{8214220} \textit{\\For 2-destination rules and sublinear $x$: Is there a $x$-round loop-free update schedule?~\cite{Forster2016Power}} & Is there a 2-round loop-free update schedule?~\cite{8214220} & In the worst case, $\Omega(n)$ rounds may be required.~\cite{8214220}, \textit{\cite{Forster2016Consistent}}. $O(n)$-round schedules always exist~\textit{\cite{roger}}. Both applies to flow-based \& \textit{destination-based} rules. \tabularnewline
\hline 
\# Rounds, relaxed LF & No results known. & 
$O(\log n)$-round update schedules always exist.~\cite{8214220} & It is not known whether $o(\log n)$-round schedules exist (in the worst case). No approximation algorithms are known.   \tabularnewline
\hline 
\# Links, strong LF & Is it possible to update $x$ nodes in a loop-free manner?~\cite{saeed-arxiv}, \textit{\cite{Forster2016Power}}  & 
Polynomial-time optimal algorithms are known to exist
in the following cases:
A maximum SLF update set can be computed in polynomial-time in trees with
two leaves.~\cite{saeed-arxiv} & The optimal SLF schedule is 2/3-approximable in polynomial time in scenarios
with exactly three leaves. For scenarios with four leaves, there exists a polynomial-time
7/12-approximation algorithm.~\cite{saeed-arxiv} Approximation algorithms from maximum acyclic subgraph~\cite{saeed-arxiv} and minimum feedback arc set~\textit{\cite{Forster2016Consistent}} apply. \tabularnewline
\hline 
\# Links, relaxed LF & Is it possible to update $x$ nodes in a loop-free manner?~\cite{saeed-arxiv} & 
Polynomial-time optimal algorithms are known to exist
in the following cases:
A maximum RLF update set can be computed in polynomial-time in trees with
two leaves.~\cite{saeed-arxiv} & No approximation results known.~\cite{saeed-arxiv} \tabularnewline
\hline 
\end{tabular}
\centering
\vskip2pt
{\textit{Note}:}{ Results/references in \textit{italics} are in the destination-based model.}
\end{table*}

Ludwig \etal in \cite{podc15}, extended in~\cite{8214220}, and \cite{hotnets14update} initiated the study of arbitrary route updates:
routes which are not necessarily destination-based.
The authors show that the update problem in this case boils down to an optimization problem on a very simple directed graph: 
initially, before the first update round, the graph simply consists of two connected paths, the old and the new route. In particular, every network node which is not part of both routes can be updated trivially, and hence,
there are only three types of nodes in this graph: the source $s$ has out-degree 2 (and in-degree 0),
the destination $d$ has in-degree 2 (and out-degree 0), and every other node has in-degree and out-degree 2, as shown in Fig.~\ref{fig:strongrelax}.
The authors also observe that loop-freedom can come in two flavors, strong and relaxed loop-freedom~\cite{8214220}.

Despite the simple underlying graph, however, Amiri \etal\cite{saeed-arxiv} show that the node-based optimization problem is NP-hard, both in the strong and the relaxed loop-free model (SLF and RLF).
In the example of Fig.~\ref{subfig:strong1} the maximization problem is easy though, clearly no more than two nodes ($s$ and $v1$) can be updated initially.
As selecting a maximum number of nodes to be updated in a given round (i.e., the node-based optimization objective)
may also be seen as a heuristic for optimizing the number of update rounds (i.e., the round-based optimization objective), the authors refer to the node-based approach as the ``greedy approach''.
Amiri \etal\cite{saeed-arxiv} also present polynomial-time optimal algorithms for specific scenarios, and both\footnote{We note that there exists a subtle difference between the approximation results by Foerster \etal and Ludwig \textit{et al.}: the former authors usually aim to minimize the number of links which \emph{cannot} be updated~\cite{Forster2016Power} (a feedback arc set problem), while Ludwig \etal\cite{saeed-arxiv} consider the dual problem variant and aim to maximize the links which \emph{can} be updated in the given round (the maximum acyclic subgraph problem). The approximation guarantees of the two problems differ: for the former model, the best known approximation bound
is $O(\log n \log\log n)$~\cite{guy-bound} while for the latter, constant approximation results exist~\cite{HASSIN1994133}.}
\cite{saeed-arxiv,Forster2016Power} provide further insights into approximability properties, see Table~\ref{loop-table}.

\smallskip
\paragraph{Round-based objective}
Ludwig \etal\cite{8214220,podc15} initiate the study of consistent network update schedules which minimize
the number of interaction rounds with the controller:
\emph{How many communication rounds $k$ are needed to update
a network in a (transiently) loop-free manner?}
The authors show that answering this question is difficult
in the strong loop-free case.
In particular, they show that while deciding whether a 
$k$-round schedule exists is
trivial for $k=2$, it is
already NP-complete for $k=3$. Moreover, the authors
show that
there exist problem instances which require $\Omega(n)$ rounds, where $n$ is the network size, see Fig.~\ref{fig:strongrelax}.
Furthermore, the authors show that 
the greedy approach, aiming to ``greedily''
update a \emph{maximum} number of nodes in each round,
may result in $\Omega(n)$-round schedules in instances which actually can be
solved in $O(1)$ rounds; even worse,
a \emph{single} greedy round may inherently delay the schedule by a factor of $\Omega(n)$ more rounds.

However, fast schedules exist for 
\emph{relaxed loop-freedom}: 
the authors present a deterministic update scheduling
algorithm which completes in $O(\log n)$-round 
in the worst case.

\smallskip
\paragraph{Other objectives}
Dudycz \etal\cite{dsn16}\rev{, detailed in~\cite{dsn16-new},} initiated the study of how to update multiple policies
simultaneously, in a loop-free manner.
In their approach, the authors aim to minimize
the number of so-called \emph{touches}: 
the total number of update messages the switches receive from the controller.
The number of touches can be reduced if controllers \emph{bundle} the updates of multiple flows to a given switch into a single message. 
However, consistency requirements impose limits on the extent to which updates can be bundled:  e.g., in order to preserve loop-freedom, the update of a flow $f_1$ needs to take place at node $v1$ \emph{before} $v2$, while  the update of a flow $f_2$ needs to take place at node $v1$ \emph{after} $v2$.
The authors establish connections to 
the \emph{Shortest Common
Supersequence (SCS)} and
\emph{Supersequence Run}
problems~\cite{superrun}, 
and show NP-hardness already for 
two policies, each of which 
can be updated in two rounds,
by a reduction from \emph{Max-2SAT}~\cite{max2sat}.

Notwithstanding, Dudycz \etal\cite{dsn16} also present optimal
polynomial-time algorithms to 
combine consistent update schedules computed for individual
policies (e.g., using any existing algorithm, e.g.,~\cite{8214220,roger}),
into a global schedule guaranteeing a minimal number of touches.
This optimal merging algorithm is not limited to loop-free updates,
but applies to any consistency property: if the consistency
property holds for individual policies, then it also holds in the joint
schedule minimizing the number of touches.

\subsubsection{Related Optimization Problems}

The link-based optimization problem, the problem of maximizing
the number of links (or equivalently nodes) which can be updated
simultaneously, is an instance of the maximum acyclic subgraph
problem; equivalently, the dual problem of minimizing the number of
links which cannot be updated is a minimum feedback arc set problem.
For showing NP-hardness, reductions from the \emph{minimum hitting set problem}~\cite{saeed-arxiv} and the \emph{feedback arc set problem}~\cite{Forster2016Power} are used. 

The problem of reconfiguring routes in a network can be seen
as a special case of \emph{combinatorial reconfiguration theory}:
an abstract reconfiguration framework to transform a feasible
solution of a problem (e.g., shortest path routing) into another solution of the same problem, e.g., while ensuring shortest path routing during the update~\cite{DBLP:journals/tcs/KaminskiMM11}).

\subsection{\rev{Guaranteeing Blackhole-Freedom}}
Another consistency property is blackhole freedom, i.e., a switch should always have a matching rule for any incoming packet, even when rules are updated (e.g., removed and replaced). This property is easy to guarantee by implementing some default matching rule which is never updated, which however could in turn induce forwarding loops. A straightforward mechanism, if there is currently no blackhole for any destination, is to install new rules with a higher priority, and then delete the old rules~\cite{Forster2016Consistent,roger}. Nonetheless, in the presence of memory limits and guaranteeing loop-freedom, finding the fastest blackhole-free update schedule is NP-hard~\cite{Forster2016Consistent}.

\rev{
\subsection{Summary and Insights}
Loop- and blackhole-freedom are both fundamental consistency properties, as their violation disconnects the logical routing graph, with loops additionally creating congestion.
Both are easy to maintain, but hard to optimize regarding makespan or resource consumption.
Of the two, loop-freedom is better understood, as already simple greedy approaches perform relatively well in simulations~\cite{Forster2016Consistent}, where the node-based objective can also be approximated well~\cite{Forster2016Power,saeed-arxiv}.
However, regarding the makespan, greedy approaches perform poorly in the adversarial scenarios~\cite{8214220}.
Still, both round- and node-based objectives are NP-hard to optimize~\cite{Forster2016Power,8214220,saeed-arxiv}.
So far, to obtain a logarithmically competitive algorithm for the number of rounds, the consistency guarantees have to be slightly relaxed~\cite{8214220}.
W}e summarize current hardness and algorithmic results in Table~\ref{loop-table}, denoting results of the destination-based model in \textit{italics}, whereas the remaining entries refer to route updates which are not necessarily destination-based.

\subsection{Open Problems}
\label{subsec:connectivity-problems}

Loop-free network updates still pose several open problems. Regarding the node-based objective,
Amiri \etal\cite{saeed-arxiv} conjecture that update problems on bounded directed path-width graphs may 
still be solvable
efficiently: none of the negative results for bounded degree graphs on graphs of bounded
directed treewidth seem to be extendable to digraphs of bounded directed pathwidth with
bounded degree.
More generally, the question of on which graph families
network update problems can be solved optimally in polynomial time
in the node-based objective remains open.
Regarding the round-based objective,
it remains an open question whether strong
loop-free updates are NP-hard for any $k\geq 3$ (but
smaller than $n$): so far only $k=3$ has been proved
to be NP-hard. More interestingly, it remains
an open question whether the relaxed loop-free
update problem is NP-hard, e.g., are 3-round update
schedules NP-hard to compute also in the relaxed
loop-free scenario? Moreover, it is not known whether
$\Omega(\log n)$ update rounds are really needed in the worst-case
in the relaxed model, or whether the problem can always
be solved in $O(1)$ rounds. Some brute-force 
computational results presented in~\cite{8214220}
indicate that if it is constant, the constant must be large.
\rev{Regarding blackhole-freedom, the possible speedup while maintaining loop-freedom is inherently connected to the available memory, but a deeper algorithmic understanding is still missing~\cite{Forster2016Consistent}.}

\section{\rev{Update Techniques to Guarantee\\Policy Consistency}}\label{sec:policies}

Modern requirements go often beyond connectivity.
For example, operators may want to ensure that packets traverse a given
middlebox (e.g., a firewall) for security reasons, or a chain of middleboxes
(e.g., encoder and decoder) for performance reasons; and/or they might like to
enforce that paths comply with Service Level Agreements (e.g., in terms of
delay). In this section, we discuss studied problems and proposed techniques
aiming at preserving such requirements during network updates.

\subsection{Definitions}
Requirements on forwarding paths additional to connectivity can be modeled by
routing \textit{policies}, that is, links, nodes or sub-paths that have to be
traversed by certain traffic flows at any moment in time.

Over the years, several contributions have targeted updates focusing on preserving specific policies.
Historically, the first policy considered during SDN updates is \textit{per-packet consistency (PPC)}, which ensures that every packet travels either on its initial or on its final paths, never on intermediate ones.
PPC seems a natural choice to comply with high-level network requirements.
Assume indeed that both the initial and the final paths accommodate requirements like security, performance, and SLA compliance.
The most straightforward way to guarantee that those requirements are not violated is to constrain all paths installed during the update to always be either initial paths or final ones.

\begin{figure}[t]
   \centering
   \subfloat[Extended Surpassed state]{
			\includegraphics[width=0.45\columnwidth]{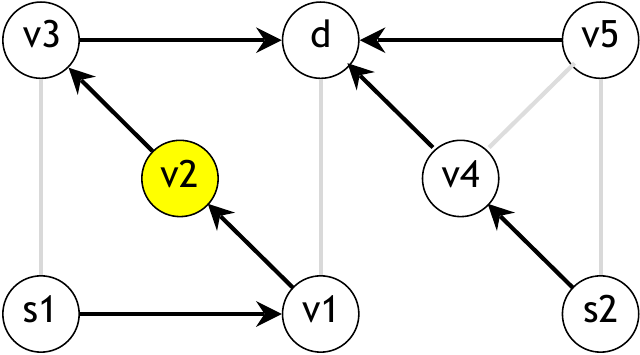}
      \label{subfig:example-wpe-init}
   }
   \hfill    \subfloat[Novel state]{
			 \includegraphics[width=0.45\columnwidth]{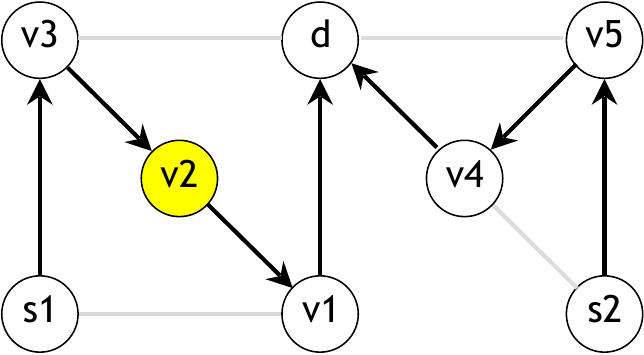}
      \label{subfig:example-wpe-fin}
   }
   \caption{A WPE-consistent update example where forwarding paths have to be changed from the Surpassed (Fig.~\ref{subfig:example-wpe-init}) to the New (Fig.~\ref{subfig:example-wpe-fin}) state, while preserving traversal of the waypoint $v2$ (highlighted in the figure) at any time during the update.}
   \label{fig:example-wpe}
\end{figure}

Nonetheless, guaranteeing PPC may be an unnecessarily strong requirement in practice.
Not always it is strictly needed that transient paths must coincide with either the initial or the final ones.
For example, in some cases (e.g., for enterprise networks), security may be the only major concern, and it may translate into simply enforcing that some flows traverse a firewall.
Fig.~\ref{fig:example-wpe} shows an example of this case, where the flow from $s1$ to $d$ has to traverse $v2$.
We refer to the property of forced traversal of a given node (waypoint) as \textit{waypoint enforcement (WPE)}.

Policies more complex than WPE (but less constraining than PPC) may also be needed in general.
For example, it may be desirable in large Internet Service Providers that specific traffic flows follow certain sub-paths (e.g., with low delay for video streaming and online gaming applications) or are explitictly denied to pass through other sub-paths (e.g., because of political or economical constraints).
Such \textit{arbitrary policies} are also considered in recent SDN update works.

\subsection{Algorithms and Complexity}\label{subsec:policies-algo}

Table~\ref{table:policies} overviews solving algorithm and complexity of
policy-preserving update problems, further discussed in the following.

\begin{table*}[htbp]
	\caption{Overview of results for policy-preserving updates.}
		\label{table:policies}
		\centering
		\scriptsize
		\begin{tabular}{|>{\centering}p{1.7cm}|>{\centering}p{3.2cm}|>{\centering}p{1.4cm}|>{\centering}p{1.2cm}|>{\centering}p{8.2cm}|}
			\hline
			Ref. & Approach & Guarantees & Computation & Remarks \tabularnewline 
			\hline  
			\hline
			\cite{jaq4,abstractions} & 2-phase commit & PPC & Constant & Always applicable (if switches have free memory slots); requires packet tagging and additional rules on internal switches \tabularnewline
			\hline
			\cite{katta2013incremental} & incremental 2-phase commit & PPC & Exponential & Always applicable (if switches have at least one free memory slot); spreads switch-memory overhead over time \tabularnewline
			\hline
			\hline
			\cite{rick1} & packet storing protocol & PPC & Constant & Dedicated protocol, based on storing packets at the controller \tabularnewline
			\hline
			\cite{rick2} & per-switch update protocol & PPC & Polynomial & Dedicated protocol, based on the translation of updates into logic circuits \tabularnewline
			\hline
			\cite{huafoum} & anti-tampering update protocol & PPC & Polynomial & Dedicated protocol, based on updating switches according to paths; Robust to tampering and dropping attacks. \tabularnewline
			\hline
			\hline
			WayUp~\cite{hotnets14update} & ordered rule replacements & WPE & Polynomial & Finishes in 4 rounds; Does not guarantee connectivity (e.g., for loops) \tabularnewline
			\hline
			MIP of~\cite{hotnets14update} & ordered rule replacements & WPE & Exponential & Optimizes update time; Guarantees absence of loops \tabularnewline
			\hline
			MIP of~\cite{sigmetrics16} & ordered rule replacements & WPE chains & Exponential & Optimizes update time; Guarantees absence of loops \tabularnewline
			\hline
			\cite{mcclurg2015efficient}, \cite[\S 2]{mcclurg} & ordered rule replacements & arbitrary & Exponential & Update synthesis based on linear temporal logic and model checking \tabularnewline
			\hline
			\cite{NateDisc16} & ordered rule replacements & PPC & Polynomial & Algorithm based on necessary conditions for PPC-preserving switch updates; Minimizes update rounds \tabularnewline
			\hline
			GPIA~\cite{safe-hybrid-upds} & ordered rule replacements & PPC & Polynomial & Greedy algorithm; Minimizes update rounds;  Applicable to hybrid SDNs (becomes exponential) \tabularnewline
			\hline
			\hline
			GPIA+FED~\cite{safe-hybrid-upds} & mixed & PPC & Polynomial & Applies restricted 2-phase commit after rule replacement sequence; Aims at reducing \# of update rounds; Applicable to hybrid SDNs (becomes exponential) \tabularnewline
			\hline
			\cite{vissicchio2016flip} & mixed & arbitrary & Exponential & Optimizes the interleaving of rule replacements and additions; Aims at reducing \# of update rounds \tabularnewline
			\hline
		\end{tabular}
	\centering
	\vskip2pt
	{\textit{Note}:}{ Contributions are grouped by approach and year.}
\end{table*}

\smallskip
\subsubsection{2-Phase commit techniques}
As described in \S\ref{subsec:taxo-taxo}, 2-phase commit techniques carry out
updates by setting an initial or final tag on packets at ingress devices (e.g.,
on $s1$ and $s2$ in Fig.~\ref{fig:example-wpe}), and maintaining two forwarding
rules at internal nodes (e.g., $v1$, $v2$, $v3$, $v4$, and $v5$ in
Fig.~\ref{fig:example-wpe}) so that each packet is forwarded over either its
initial or final paths, according to the carried tag.
This approach guarantees per-packet consistency by design.

A framework to implement 2-phase commit in traditional networks has been
proposed by Alimi \etal\cite{alimi_shadow_config_2008} (see also \S\ref{sec:history}).
It requires invasive modification of router internals, to manage tags and run
arbitrary routing processes in separate process spaces.
Such modifications are not needed in SDNs, where data-plane devices exhibit
finer-grained programmability.
Beyond presenting an implementation of 2-phase commit in OpenFlow, pioneering
SDN update works~\cite{jaq4,abstractions} argue for the criticality of ensuring
PPC in the SDN case, and of providing programmatic support for consistent
updates within SDN controllers.

A downside of 2-phase commit techniques is that they require internal switches
(e.g., $v1$, $v2$, $v3$, and $v5$ in Fig.~\ref{fig:example-wpe})
to maintain an additional rule every time their initial and final rules differ.
This overhead requires all switches to have free memory slots (whose number
depends on the update case), generally wastes memory resources, and might hamper
other applications to properly work during updates (e.g., fast rerouting or
security ones that might need to also install new flow rules in reaction to
sudden traffic changes).
To mitigate those problems, an incremental version of the original technique has
been studied in~\cite{katta2013incremental}.
This work proposes to divide the input update into sub-updates that can be
carried out one after the other.
Consider for example a variant of Fig.~\ref{fig:example-wpe} where $d$ is
replaced by a set $\cal D$ of $N$ replicas $d_1,\dots,d_N$, and the update
consists in changing the paths for every $d_i \in {\cal D}$ as shown in the
figure. The incremental 2-phase commit technique enables to break down such an
update into a sequence of sub-updates, where each sub-update modifies the paths
of a distinct set of destinations ${\cal D}_j \subset \cal D$.
This will limit the rules added to every switch at any time, 
at the price of a longer update completion time.

Ultimately, switch-memory consumption remains a fundamental limitation of of
2-phase commit techniques, since tag-matching rules must be added to internal
switches sooner or later during an update (or sub-updates).
Both the original and the incremental techniques also exhibit other limitations,
like the need for packet-header space, the tagging overhead, and complications
with middleboxes modifying packet headers and
tags~\cite{DBLP:conf/sigcomm/FayazbakhshSYM13,DBLP:conf/sigcomm/QaziTCMSY13}.

\smallskip
\subsubsection{SDN-based update protocols}
McGeer~\cite{rick1,rick2} presented two protocols to carry out network updates defined on top of OpenFlow.
The first update protocol~\cite{rick1} is based on sending packets to the
controller during updates, so that the controller can locally store packets
until the final flow rules are installed on all the switches. As a result,
switch resources (especially, TCAM entries) are saved, at the cost of adding
delay on packet delivery, consuming network bandwidth, and requiring the
controller to temporarily store packets.
The second protocol~\cite{rick2} implements sequences of per-switch rule updates
that guarantee PPC: It updates one switch at the time, ensuring that PPC
is preserved at every step.
In addition, Hua \etal\cite{huafoum} initiate a study on how to support PPC in an
adversarial setting. They present FOUM, an update protocol where the update
sequence is encoded in a packet signed by the controller, sent to one switch,
and passed among switches at runtime. FOUM is shown to be robust to
packet-tampering and packet dropping attacks.
All those works assume dedicated protocols that are not supported by devices out
of the box.

\begin{figure*}[t]
	\centering
	\subfloat[Step 1]{
		\includegraphics[width=0.45\columnwidth]{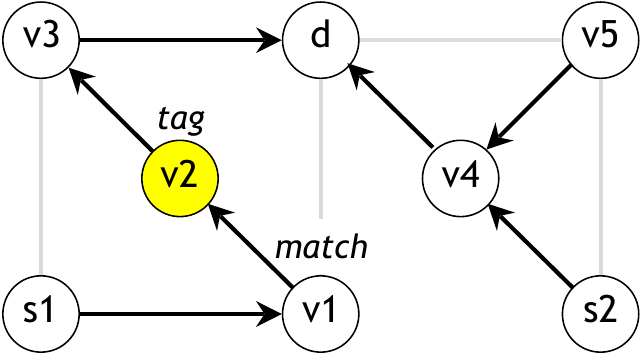}
		\label{subfig:flip-seq1}
	}\quad
	\subfloat[Step 2]{
		\includegraphics[width=0.45\columnwidth]{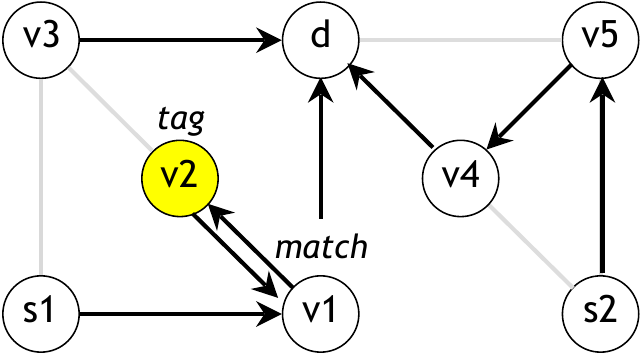}
		\label{subfig:flip-seq2}
	}\quad	\subfloat[Step 3]{
		\includegraphics[width=0.45\columnwidth]{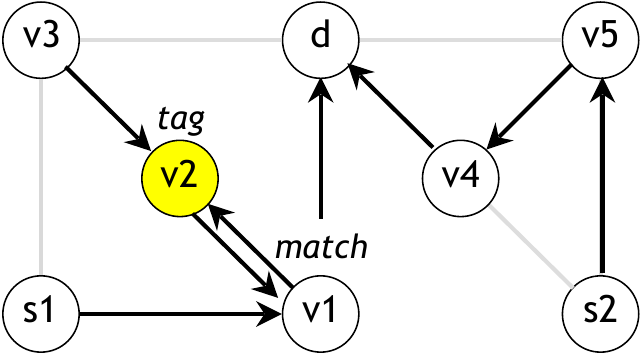}
		\label{subfig:flip-seq3}
	}\quad
	\subfloat[Step 4]{
		\includegraphics[width=0.45\columnwidth]{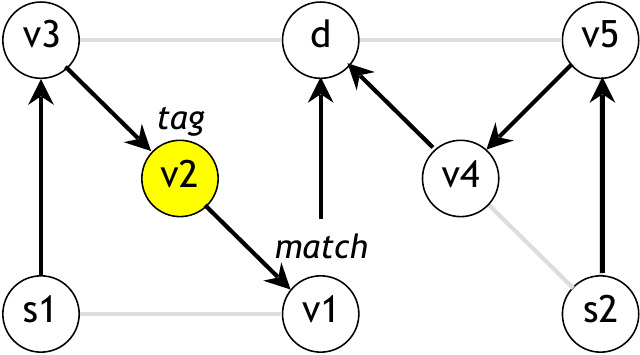}
		\label{subfig:flip-seq4}
	}
	\caption{Sequence generated by the FLIP algorithm proposed in~\cite{vissicchio2016flip}. Note that the loop between $v1$ and $v2$ in Step 2 and 3 does not break connectivity, as it is traversed \textit{only once} by packets. This is because $v1$ matches the tag set by $v2$, hence it forwards packets that have already traversed $v2$ directly to $d$.}
	\label{fig:flip-sequence}
\end{figure*}

\smallskip
\subsubsection{Rule replacement ordering}
Further works explore which policies can be supported by only relying on
carefully-computed sequences of rule replacements, so as to (i) introduce
no memory overhead, and (ii) be readily supported by all (traditional
and SDN) devices, without tag-related issues.

Initial contributions mainly focused on WPE consistency.
Prominently,~\cite{hotnets14update} studies how to compute short sequences of
rule replacements, for quick updates ensuring that any given flow traverse a
single waypoint. The authors propose WayUp, an algorithm that computes
WPE-preserving updates spanning 4 rounds. However, they also show that it is not
possible to guarantee both WPE and loop-freedom in some cases.
Fig.~\ref{fig:example-wpe} actually shows one case in which any sequence of rule
replacements either causes a loop or a WPE consistency violation. Those
infeasibility results have been extended to chains of waypoints
in~\cite{sigmetrics16}. The latter work shows that flexibility in ordering and
placing virtualized functions specified by a chain do not make the update
problem always solvable.
Those two works also prove that it is NP-hard to even decide if there exists a
sequence of rule replacements preserving both loop-freedom and WPE (or waypoint
chain traversal). Mixed Integer Program (MIP) formulations to find a safe
sequence of rule replacements (when any exists) are proposed and evaluated in
both cases.

The more general problem of preserving arbitrary policies defined by
operators is tackled in~\cite{mcclurg2015efficient}. This paper describes an
approach to (i) model update-consistency properties as Linear Temporal Logical
formulas, and (ii) automatically synthesize SDN updates that preserve input
properties. Such a synthesis is performed by an algorithm based on
counterexample-guided search and incremental model checking. Experimental
results are provided about the scalability of the algorithm (up to networks with
1,000 nodes).

More recent works finally consider the problem of guaranteeing PPC by ordering
rule replacements.
Vissicchio \etal\cite{safe-hybrid-upds} show that this problem can be solved efficiently;
they prove that a polynomial-time greedy algorithm called GPIA finds the sequence
of per-switch rule replacements that does not violate PPC while updating the
maximal number of switches and allowing the maximal parallelism between
per-switch updates. The algorithm is based on iteratively simulating the update
of every switch which has not been updated yet.
For instance, in Fig.~\ref{fig:example-wpe}, GPIA would
initially simulate the replacement of next-hops on every node.
It would then define the first round of the update under computation by
collecting all the switches ($v4$ and $v5$ in this example) whose update does
not violate PPC.
Then, it would iterate on the remaining nodes, discovering that $s2$ can be
updated in the second round without violating PPC.
GPIA terminates when there are no switches that can be safely updated.
Cern{\'{y}} \etal\cite{NateDisc16} describe a refined version of this algorithm that avoids the
simulation of switch updates thanks to the identification of necessary
conditions for safely updating the switches.

Opportunities and limitations of rule replacement ordering for PPC-preserving
updates have also been evaluated in~\cite{safe-hybrid-upds}, by simulating
realistic update scenarios, on real network topologies.
Results show that ordered rule replacements can rarely complete an update (for
example, it could not in Fig.~\ref{fig:example-wpe}, but they can safely update
many switches (typically, even more than the $3$ out of $7$ that it would update
in Fig.~\ref{fig:example-wpe}\rev{)}).
Those results further motivate rule-replacement algorithms tailored to a more
restricted family of policies (like WPE-preserving ones) as
well as mixed approaches (employing both rule replacements and duplication,
as described below).

\smallskip
\subsubsection{Mixed approaches}
Following up on their experimental results, Vissicchio \etal\cite{safe-hybrid-upds} argue for
sequentially computing a safe rule replacement ordering, and applying a
scope-limited 2-phase commit variant afterwards.
In the example of Fig.~\ref{fig:example-wpe}, this approach would then lead to
concatenating the ordered rule replacements on $v4$, $v5$ and $s2$ with the
application of a 2-phase commit technique to the sub-network of non-updated
switches (i.e., $s1$, $v1$, $v2$ and $v3$).
This combination ensures the possibility to always perform the update (contrary
to pure rule-replacement approaches) while reducing the update overhead (in
switch memory and data-plane) with respect to the original 2-phase commit
technique.
The same work also generalizes this strategy to hybrid SDNs, potentially running
any number both traditional and/or SDN control-planes -- a setting in which a
brute force (exponential) algorithm might be needed, depending on the nature of
control-planes involved in the update.

FLIP, a different algorithm implementing a mixed approach, is described
in~\cite{vissicchio2016flip} and detailed in~\cite{flip-ton17}.
FLIP jointly optimizes the interleaving of rule replacements and additions (for
matching packet tags) so as to preserve arbitrary policies, including PPC.
For instance, in Fig.~\ref{fig:example-wpe}, FLIP would compute a sequence of
operations (illustrated in Fig.~\ref{fig:flip-sequence}) where only $v1$ matches
a tag set by $v2$: That is, FLIP needs only 1 additional rule when 2-phase
commit would add 4, and sequentially combining replacements and additions would
result in 3 additional rules.
This implies that FLIP is strictly more powerful (i.e., solves a
higher number of update cases) than only relying on rule replacements,
exclusively using 2-phase commit, and sequential combining those two approaches.
However, it is unclear how FLIP can be used in hybrid SDNs, with more than
one control-plane.
Also, its time complexity is not polynomial -- even if the experiments suggest
that FLIP quickly computes short update sequences in realistic networks.

\subsection{Related Optimization Problems}
Many policy-preserving algorithms face generalized versions of the optimization
problems associated to connectivity-preserving updates (see
\S\ref{sec:forwarding}): While the most common objective remains the
maximization of parallel operations (to speed-up the update), policy consistency
requires that all possible intermediate paths comply with certain regular
expressions in addition to being simple (that is, loop-free) paths.
Mixed policy-preserving approaches focus on even more general problems where
(i)~ different operations can be interleaved in the output operational sequence
(which provides more degrees of freedom in solving the input problems), and
(ii)~ multiple optimization objectives are considered at the same time
(typically, maximizing the update parallelism while also minimizing the consumed
switch memory).

\subsection{\rev{Summary and Insights}}
Unsurprisingly, preserving policies requires more sophisticated update
techniques, since it is generally harder to extract policy-induced constraints
and model the search space.
Two major families of solutions have been explored so far.
On the one hand, 2-phase commit techniques and update protocols sidestep the
algorithmic challenges, at the cost of relying on specific primitives (packet
tagging and tag matching) that comes with switch memory consumption.
On the other hand, ordering-based techniques directly deal with problem
complexities, at the cost of algorithmic simplicity and impossibility to always
solve update problems.
\rev{Initial work has been done on mixed approaches, relying on algorithms that can
interleave different kinds of operations within the computed operational
sequence.}

\subsection{Open Problems}
\label{subsec:policy-problems}

Finding the best balance between the two extremes of relying on protocols on one
hand, and on ordering algorithms on the other hand is an interesting direction.
Despite some initial work has started towards this goal (see
\S\ref{subsec:policies-algo}), many research questions are left open.
For example, the computational complexity of solving update problems while
mixing rule additions (for packet tagging and matching) with replacements is
unknown.
Moreover, it is unclear whether the proposed algorithms can be improved
exploiting the structure of specific topologies or the flexibility of new
devices (like those implementing P4~\cite{p4-ccr14}) -- for example, to achieve
a better trade-off between switch memory consumption and update speed.

\section{\rev{Update Techniques to Guarantee\\Congestion-Aware Consistency}}\label{sec:cap}

\begin{table*}[hbtp]
\caption{Compact overview of flow migration algorithms.} \label{flow-algo-table}
\centering
\scriptsize

\begin{tabular}{|>{\centering}p{1.6cm}|p{2.5cm}|>{\centering}p{1.7cm}|>{\centering}p{0.51cm}|>{\centering}p{1.3cm}|p{1.16cm}|p{5cm}|}
\hline
 Ref. & Approach & (Un-)splittable model & Interm. paths  & Computation & \# Updates & Complete (decides if consistent migration exists)  \tabularnewline 
\hline  
\hline
\cite{abstractions} & Install old and new rules, then switch from old to new & Both, move each flow only once & No  & Polynomial & 1 & No bandwidth guarantees  \tabularnewline
\hline 
\cite{swan}& Partial moves according to free slack capacity $s$ & Splittable & No  & Polynomial & $\left\lceil 1/s\right\rceil-1$ & Requires slack on flow links \\ \hline 
~\cite{dionysus}& Greedy traversal of dependency graph & Both, move each flow only once  & No & Polynomial & Linear & No (rate-limit flows to guarantee completion) \\ 
\hline
\cite{DBLP:conf/icc/LuoYLZ15}& MIP of ~\cite{dionysus} & Both, move each flow only once  & No & Exponential & Linear & Yes \\ 
\hline

\multirow{3}{*}{}\cite{icnp-jiaqi} & \multirow{3}{2.5cm}{Fix number of $x$ inter- mediate states ahead of time, optimize via LP
} & \multirow{3}{*}{}Both & No & Polynomial & \multirow{3}{*}{}Any $x \in \mathbb{N}$ & \multirow{3}{6cm}{For a given number of intermediate states $x$, approximate minimum transient congestion (if $>0$) by $\log n$ factor} \tabularnewline
\cline{4-5} 
 &  &  & Yes & \emph{Exponential} &  &   \tabularnewline
 &  &  &  &  &  &   \tabularnewline
\hline 
\cite{icnp-jiaqi}& ... via MIP & Both  & Both & \emph{Exponential} & Any $x \in \mathbb{N}$  & For any given $x$ yes, but not in general \\ \hline
~\cite{swan}& Binary  search of intermediate states via LP & Splittable & Yes  & Polynomial in \# of updates & Unbounded & Cannot decide if consistent migration is possible  \\  \hline 

~\cite{roger-infocom}& Create slack with intermediate states, then use partial moves of~\cite{swan} & Splittable  & Yes & Polynomial &  Unbounded & Yes \tabularnewline
\hline 

\cite{Forster2016Power}& Split flows along old and new paths & 2-Splittable & No & Polynomial & Unbounded  & Yes \\
\hline
\cite{SebastianBrandt2016}& Use augmenting flows to find updates & Split., 1 dest., paths not fixed  & Yes & Polynomial & Linear & Yes\\
\hline
\cite{DBLP:conf/nca/FoersterJSW16}& Dynamic programming & Unsplittable  & Both & \emph{Exponential} & Exponential & Yes\\
\hline

\hline
 \multicolumn{7}{|l|}{Further practical extensions} \tabularnewline
\hline
~\cite{zupdate} & \multicolumn{6}{|l|}{Extends approach of \emph{SWAN}~\cite{swan} in a data center setting}   \tabularnewline
\hline 
~\cite{WangHeSuEtAl2016} & \multicolumn{6}{|l|}{Extends approach of \emph{Dionysus}~\cite{dionysus} with local dependency resolving}   \tabularnewline
\hline 
~\cite{DBLP:conf/sigcomm/JinLWLGXLXR16} & \multicolumn{6}{|l|}{Extends approach of \emph{Dionysus}~\cite{dionysus} with circuit nodes for optical wavelengths}   \tabularnewline
\hline 
~\cite{DBLP:conf/icc/ChenW17} & \multicolumn{6}{|l|}{Extends approach of \emph{Dionysus}~\cite{dionysus} by using switch buffers to break deadlocks}   \tabularnewline
\hline 
~\cite{DBLP:conf/conext/GandhiRJ17} & \multicolumn{6}{|l|}{Extends approach of \emph{Dionysus}~\cite{dionysus} by allowing multiple target states}   \tabularnewline
\hline 
~\cite{ParisDestounisMaggiEtAl2016} & \multicolumn{6}{|l|}{Considers reconfiguration for dynamic flow arrivals}   \tabularnewline
\hline 
~\cite{LuoYuLuoEtAl2016} & \multicolumn{6}{|l|}{Allows (un-)splittable flow migration (move once) with user-spec. deadlines \& requirements via  MIP (LP heuristic)}   \tabularnewline
\hline 
\rev{\hspace{-0.15cm}\cite{DBLP:conf/nca/DelaetDKTG15,DBLP:conf/nca/DinitzDK17,DELAET201881}} & \multicolumn{6}{|l|}{\rev{Proposes multi-casting on portions of routes for faster updates, also to break deadlocks by using out-of-band capacities}}   \tabularnewline
\hline 
\multirow{1}{*}{\parbox{2cm}{\hspace{-0.15cm}\cite{DBLP:journals/corr/AmiriDSW16}, \rev{\cite{DBLP:conf/icalp/AmiriDSW18,DBLP:journals/corr/abs-1805-06315}}}} & \multicolumn{6}{|l|}{\makecell[l]{Does not require tagging of flows in the packet header, flows may take a mix of the old and new paths. For a constant number of flows on directed\\ acyclic graphs (DAGs), a linear-time (fixed parameter tractable) algorithm is provided.}}   \tabularnewline
\hline 

\end{tabular}
\end{table*}

Computer networks are inherently capacitated, and respecting resource constraints is hence another important aspect of consistent network updates.  
Congestion is known to significantly impact throughput and increase latency, therefore negatively impacting user experience and even leading to unpredictable economic loss. 

\subsection{Definitions}
The capacitated update problem is to migrate from a multi-commodity flow $\mathcal F_{old}$ to another multi-commodity flow $\mathcal F_{new}$, where consistency is defined as not violating any link capacities and not rate-limiting any flow below its demand in $\min\left(\mathcal F_{old}, \mathcal F_{new}\right)$. In few works, e.g.,~\cite{SebastianBrandt2016}, $\mathcal F_{new}$ is only implicitly specified by its demands, but not by the actual flow paths.
Some migration algorithms will violate consistency properties to guarantee completion, as a consistent migration does not have to exist in all cases.
 It can then be useful to investigate, e.g., what type of rate-limiting is performed~\cite{Zheng2018}.

Typically, four different variants are studied in the literature: First, individual flows may either only take one path (unsplittable) or they may follow classical flow-theory, where the incoming flow at a switch must equal its outgoing flow (splittable). Secondly, flows can take any paths via helper rules in the network during the migration (intermediate paths), or may only be routed along the old or the new paths (no intermediate paths).

To exactly pinpoint congestion-freedom, one would need to take many detailed properties into account, e.g., buffer sizes and ASIC computation times. 
As such, the standard consistency model does not take this fine-grained approach, but rather aims at avoiding ongoing bandwidth violations and takes a mathematical flow-theory point of view.
Introduced by~\cite{swan}, consistent flow migration is captured in the following model: No matter if a flow is using the rules before the update or after the update, the sum of all flow sizes must respect the link capacity.

\begin{figure*}[ht]
   \centering
   \subfloat[Initial]{
      \includegraphics[width=0.3\columnwidth]{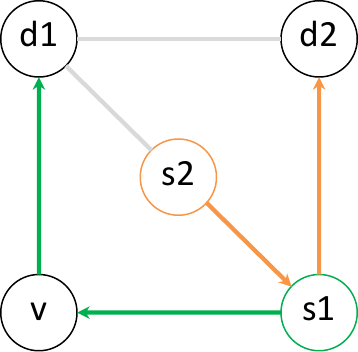}
      \label{subfig:flow-cong-ex-1}
   }\quad
   \subfloat[Congestion!]{
      \includegraphics[width=0.3\columnwidth]{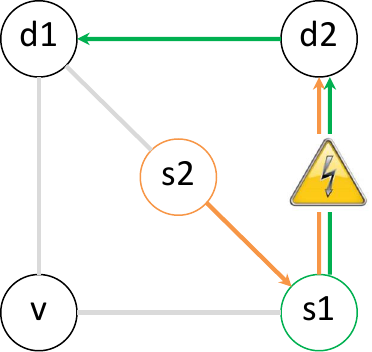}
      \label{subfig:flow-cong-ex-2}
   }\quad
   \subfloat[Orange first]{
        \includegraphics[width=0.3\columnwidth]{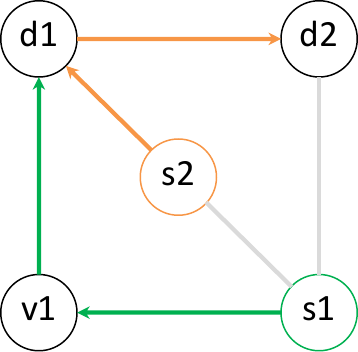}
        \label{subfig:flow-cong-ex-3}
   }\quad
   \subfloat[Final]{
        \includegraphics[width=0.3\columnwidth]{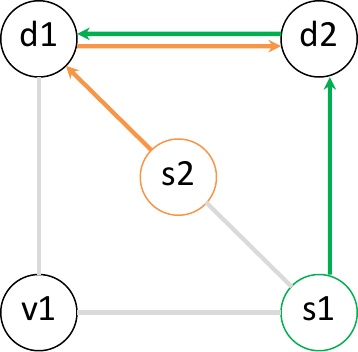}
        \label{subfig:flow-cong-ex-4}
   }
   \caption{In this flow migration example, all links have unit bidirectional capacity, and both orange and green flows have unit size as well. The task is to move both the green and orange flows from their initial paths in Fig.~\ref{subfig:flow-cong-ex-1} to their final ones shown in Fig.~\ref{subfig:flow-cong-ex-4}. Updating both flows together could lead to the green flow being moved first, inducing congestion, see Fig.~\ref{subfig:flow-cong-ex-2}. However, this can be avoided by using succinct updates, first moving the orange flow as in Fig.~\ref{subfig:flow-cong-ex-3}, then the green flow.}
   \label{fig:move up}
	
\end{figure*}

\subsection{Algorithms}

\rev{Most} current algorithms for capacitated updates of network flows use the seminal work by Reitblatt \etal\cite{abstractions} as an update mechanism. Analogously to \emph{per-packet consistency} (cf.~\S\ref{sec:policies}), one can achieve \emph{per-flow consistency} by a 2-phase commit protocol.
While this technique avoids many congestion problems, is not sufficient for bandwidth guarantees: When updating the
flows in Fig.~\ref{fig:move up}, if the green flow moves up before orange flow is on its new path, congestion occurs.

Mizrahi and Moses~\cite{DBLP:conf/ons/MizrahiM14} prove that flow swapping is necessary for throughput optimization in the general case, as thus algorithms are needed that do not violate any capacity constraints during the network update, beyond simple flow swapping as well.

An overview of current algorithmic approaches can be found in Table~\ref{flow-algo-table}.
In particular, we briefly describe the technique used, e.g., partial moves using slack capacity or dependency graphs. Furthermore, we categorize the algorithm techniques according to their model assumptions (splittability, helper rules), their complexity (computation and \# updates), and if they can decide the underlying decision problem under their model assumptions, see also the later Table~\ref{hardness results-table}.
Note that small changes in the model can lead to different complexities. For example, by not allowing intermediate paths as in, e.g.,~\cite{dionysus,DBLP:conf/icc/LuoYLZ15}, flow migration is easier to handle, but is also less powerful. In the following, we focus on selected works that introduce general techniques or model ideas.

\smallskip

\subsubsection{Slack}
\begin{figure}[thbtp]
   \centering
   \subfloat[Initial]{
      \includegraphics[width=0.3\columnwidth]{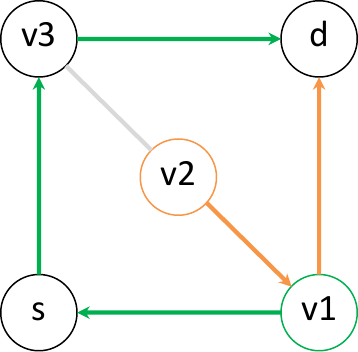}
      \label{subfig:flow-cong-swap-1}
   }
   \qquad \qquad
   \subfloat[Final]{
      \includegraphics[width=0.3\columnwidth]{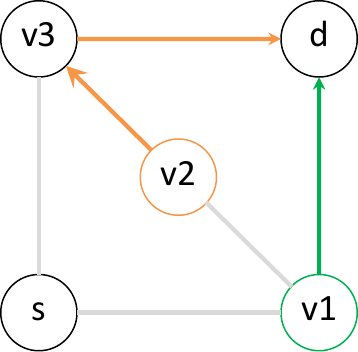}
      \label{subfig:flow-cong-swap-2}
   }
   \caption{In this network the task is to migrate consistently from the initial to the final state. If all flows and links have unit size, no consistent migration is possible: the destination has just incoming links of size two.  If the flows just have a size of $2/3$, one can migrate consistently in $\left\lceil 1/(1/3)\right\rceil-1=2$ updates by moving half of the flow size of $1/3$ each time in parallel.
   }
   \label{fig:move up-swan}
\end{figure}

 The seminal work by Hong \etal\cite{swan} on \emph{SWAN} introduces the current standard model for capacitated updates. Their algorithmic contribution is two-fold, 
and also forms the basis for~\emph{zUpdate}~\cite{zupdate}: 
First, the authors show that if all flow links 
have free capacity \emph{slack} $s$, 
consistent migration is possible using 
$\left\lceil 1/s\right\rceil-1$ updates: E.g., if the free capacity is 10\%, 
 9 updates are required, always moving 10\% of the links' 
 capacity to the new flow paths. 
If the network contains non-critical background traffic, free capacity can be generated for a migration by rate-limiting this background traffic temporarily, see~Fig.~\ref{fig:move up-swan}: removing some background traffic allows  consistent migration.

\smallskip
\subsubsection{LP-formulation}
Second, Hong \etal\cite{swan} present an LP-formulation for 
splittable flows which provides a consistent migration 
schedule with $x$ updates, if one exists. By performing a binary search over the number of updates, the number of necessary updates can be minimized. This approach allows for intermediate paths, where the flows can be re-routed anywhere in the network.
E.g., consider the example in  Fig.~\ref{fig:move up-swan} with all flows and links having unit size. If there was an additional third route to $d$, 
the orange flow could temporarily use 
this intermediate path: we can then switch 
the green flow, and eventually 
the orange flow could be moved to its desired new path.

\begin{table*}[htbp]
\caption{Table summarizing decision problem results for flow migration.}
\label{hardness results-table}
\centering
\scriptsize

\begin{tabular}{|>{\centering}p{3cm}|>{\centering}p{2cm}|>{\centering}p{2cm}|>{\centering}p{8cm}|}
\hline
Flow migration problem & Intermediate paths & Memory  restrictions & Decision problem hardness  \tabularnewline 
\hline  
\hline
\multirow{4}{*}{Unsplittable} & \multirow{2}{*}{Yes} & Yes & NP-hard~\cite{roger-infocom},\tabularnewline \cline{3-3} 

 &  & No &  EXPTIME~\cite{DBLP:conf/nca/FoersterJSW16} \tabularnewline \cline{2-4}
 & \multirow{2}{*}{No}  & Yes  & NP-hard~\cite{Forster2016Power}, \tabularnewline \cline{3-3}
 &  & No  & EXPTIME~\cite{DBLP:conf/nca/FoersterJSW16} \tabularnewline \cline{3-4}
\hline 

\multirow{4}{*}{Unit size} & \multirow{2}{*}{Yes}& Yes& NP-hard~\cite{roger-infocom},\tabularnewline \cline{3-3} 

 &  & No & EXPTIME~\cite{DBLP:conf/nca/FoersterJSW16} \tabularnewline \cline{2-4}
 & \multirow{2}{*}{No}  & Yes  &  \multirow{2}{*}{Open (also for integer size splitting)} \tabularnewline \cline{3-3}
 &  & No  &  \tabularnewline \cline{3-4}
\hline 

\multirow{4}{*}{Splittable} & \multirow{2}{*}{Yes}& Yes& NP-hard~\cite{dionysus}\tabularnewline \cline{3-4} 

 &  & No & P~\cite{roger-infocom} \tabularnewline \cline{2-4}
 & \multirow{2}{*}{No}  & Yes  & NP-hard~\cite{dionysus} \tabularnewline \cline{3-4}
 &  & No  & Open \tabularnewline \cline{3-4} \hline 
\hline

\multirow{4}{*}{\parbox{4cm}{Move every flow only once}} & \multirow{2}{*}{Yes}& Yes& \multirow{2}{*}{Not allowed (model)} \tabularnewline \cline{3-3} 

 &  & No &  \tabularnewline \cline{2-4}
 & \multirow{2}{*}{No}  & Yes  &  NP-complete~\cite{dionysus} \tabularnewline \cline{3-4}
 &  & No  & NP-complete~\cite{Forster2016Power} \tabularnewline \cline{3-4}
\hline
\hline 

\multirow{2}{*}{Node-ordering} & \multirow{2}{*}{Mix of old and new}& Yes& \multirow{2}{*}{NP-hard~\cite{DBLP:journals/corr/AmiriDSW16}} \tabularnewline \cline{3-3} 

 &  & No &  \tabularnewline \hline

\end{tabular}
\centering
\vskip2pt
{\textit{Note}:}{ In general, it is unknown if flow migration is in NP if flows can be moved more than once, except for the case of splittable flows without memory restrictions. We note that if a problem is NP-hard without memory restrictions, it is also NP-hard with memory restrictions, as providing sufficient memory is a special case of memory restrictions.
\vspace{-0.3cm}
}

\end{table*}

\smallskip
\subsubsection{Spread flows over the network}
Brandt \etal\cite{roger-infocom} prove that splittable flow migration is always decidable in polynomial time, by providing an algorithm that attempts to create slack capacity on all links. The fundamental idea is to keep splitting flows along new paths, until slack is obtained such that the algorithm of Hong \etal\cite{swan} is applicable. For example in Fig.~\ref{subfig:flow-cong-ex-1}, one can proceed as follows: 1) split the orange flow equally along the old and new path, and afterwards 2) route a quarter of the green flow via either $s2$ or $d2$.  The correctness of their approach relies on an augmenting flow techniques, we refer the reader to \cite{roger-infocom} for the intricate details.

\smallskip
\subsubsection{Dynamicity \& dependency graphs}
Jin \etal\cite{dionysus} also consider the variable update times of switches in the network. To this end, inspired by~\cite{roger}, they build a dependency graph of the individual updates, greedily sending out updates once the respective pre-conditions are satisfied. For example, assume a flow $f_1$ can be migrated to its new path when least one of the flows $f_2,f_3$ was moved, but $f_2,f_3$ take different (unknown) time-spans to complete the move: the fastest method is to dynamically wait until either $f_2,f_3$ moves, any pre-computed schedule will have a longer makespan under adversarial conditions.
When this greedy traversal of the dependency graph results in a deadlock, flows are rate-limited to guarantee progress. 
We note that it is not clear how to extend the dependency graph idea to intermediate paths.

\smallskip
\subsubsection{Jointly optimize migration \& new paths}
Brandt \etal\cite{SebastianBrandt2016} introduce the idea that the new flow paths should not be part of the problem input, but rather be computed jointly with the migration schedule. This idea aims at 1) speeding up the migration process and 2) allowing more problem instances to be solved. As an illustration, consider the problem in Fig.~\ref{fig:move up-swan}: from a pure admission perspective, no migration is needed, both flow demands are already satisfied in the initial state.
Interestingly, for a single destination (but multiple commodities), consistent migration is \emph{always} possible in this model, as long as there is some way to admit all flows without violating capacities.
A framework for general multi-commodity flows is still missing.

Ghandi \etal\cite{DBLP:conf/conext/GandhiRJ17} also observe that flows can have multiple migration options, as networks are often built with redundancy in mind. 
To this end, they first compute multiple choices for new flow paths, optimizing
for close-to-optimal path properties and few stages in the resulting dependency graph. This allows them to dynamically speed up the execution of the consistent network updates, depending on the runtime conditions.

\begin{table*}[hbtp]
\caption{Compact overview of flow migration hardness techniques and results.} \label{flow-hardness-table}
\centering
\scriptsize
\begin{tabular}{|>{\centering}p{1.7cm}|>{\centering}p{1.2cm}|>{\centering}p{1.5cm}|>{\centering}p{1.1cm}|>{\centering}p{0.7cm}|>{\centering}p{1.5cm}|p{6cm}|}
\hline
 Ref. & Reduction via & (Un-)splittable model & Interm. paths  & Memory limits & Decision prob. in general & Optimization problems/remarks  \tabularnewline 
\hline 
\hline
~\cite{dionysus} & Partition & Splittable & No & Yes & NP-hard & NP-complete if every flow may only move once\\  \hline
~\cite{dionysus} & Partition & Splittable & No & No & -- & NP-hard (fewest rule modifications)\\  \hline
~\cite{roger-infocom} & -- & Splittable & Yes & No & P & Fastest schedule can be of unbounded length, LP for new reachable demands if cannot migrate \\  \hline
~\cite{Forster2016Power} & -- & 2-Splittable & No & No & P & studies slightly different model \\  \hline
~\cite{roger-infocom} & (MAX) 3-SAT & Unsplittable & Yes & No & NP-hard (also for unit size) & NP-hard to approx. additive error of flow removal for consistency better than $7/8+\varepsilon$\\  \hline
~\cite{icnp-jiaqi} & Partition & Unsplittable & Yes/No & No & -- & NP-hard (fastest schedule)\\  \hline
~\cite{Forster2016Power} & Partition & Unsplittable & No & No & NP-hard & Stronger consistency model, but proof carries over  \\  \hline
~\cite{LuoYuLuoEtAl2016} & Partition \& Subset sum & Unsplittable & No & No & -- & NP-hard for 3-update schedule \\  \hline
~\cite{DBLP:conf/nca/FoersterJSW16} & Disjoint paths & Unsplittable & Yes & No &  NP-hard (also for unit size) & NP-hard for already 2 unit size flows \\  \hline
\hline
\parbox{2cm}{\hspace{-0.12cm}\cite{DBLP:journals/corr/AmiriDSW16}, \rev{\cite{DBLP:conf/icalp/AmiriDSW18,DBLP:journals/corr/abs-1805-06315}}} & 3-SAT & Node-ordering & Mix of old \& new & No &  NP-hard (also for unit size) & NP-hard \rev{for 6 unit size flows if the pair of old and new path forms a DAG, on general graphs 2 flows suffice}\\  \hline
\end{tabular}
\end{table*}

\smallskip
\subsubsection{Node-ordering instead of 2-phase commit}
Amiri \etal\cite{DBLP:journals/corr/AmiriDSW16}, \rev{\cite{DBLP:conf/icalp/AmiriDSW18,DBLP:journals/corr/abs-1805-06315}} propose to identify flows only by their source and destination, removing flow version numbers from the packet header (``tagging''). Their approach reduces complexity overhead, but does not permit the use of 2-phase commit techniques. Conceptually, each node has an old and a new forwarding rule for each flow, where the challenge is how to \emph{order} these updates, without inducing congestion or forwarding loops. 
For an intuition, recall Fig.~\ref{subfig:example-history-sitn-seq4}, and let the old rules be marked in solid blue, with the new rules being in dash-dotted green: over multiple rounds, the routing rules converge to the new state.

\subsection{Complexity}
The complexity of capacitated updates can roughly be summarized as follows: Problems involving splittable flows can be decided in polynomial time, while restrictions such as unsplittable flows or memory limits turn the problem NP-hard, see Table~\ref{hardness results-table}. For unsplittable flows, an exponential time algorithm exists.
In a way, the capacitated update problems differs from related network update problems in that it is not always solvable in a consistent way. On the other hand, e.g., per-packet/flow consistency can always be maintained by a 2-phase commit, and loop-free updates for a single destination can always be performed in a linear number of updates.

One standard approach in recent work for flow migration is linear (splittable flows) or integer programing (unsplittable flows): With the number of intermediate configurations $x$ as an input, it is checked if a consistent migration with $x$ intermediate states exists. Should the answer be yes, then one can use a binary search over $x$ to find the fastest schedule. 
This idea originated in \emph{SWAN}~\cite{swan} for splittable flows, and was later extended to other models, cf.~Table~\ref{flow-algo-table}.

However, the LP-approach via binary search (likewise for the integer one) suffers from the drawback that it is only complete if the model is restricted:
If $x$ is unbounded, then one can only decide whether a migration with $x$ updates exists, but not whether there is no migration schedule with $y$ steps, for some $y>x$.
Additionally, it is not even clear to what complexity class the general capacitated update problem belongs to, cf.~the decision problem hardness column of Table~\ref{hardness results-table}.

The only exception arises in case of 
splittable flows without memory restrictions, where either an (implicit) schedule or a certificate that no consistent migration is possible, is found in polynomial time~\cite{roger-infocom}. The authors use 
a combinatorial approach not relying on linear programming. Adding memory restrictions turns this problem NP-hard as well~\cite{dionysus}.

If the model is restricted to allow every flow only to be moved once (from the old path to the new path), then the capacitated update problem becomes NP-complete~\cite{Forster2016Power,dionysus}: Essentially, as the number of updates is limited by the number of flows, the problem is in NP.
In this specific case, one can also approximate the minimum congestion for unsplittable flows in polynomial time by randomized rounding~\cite{icnp-jiaqi}.

Hardly any (in-)approximability results 
exist today, and 
most work relies on reductions from the Partition problem, cf.~Table~\ref{flow-hardness-table}.
The only result that we are aware of is via a reduction from MAX 3-SAT, which also applies to unit size flows~\cite{roger-infocom}.

\subsection{Related Optimization Problems}

In a practical setting, splitting flows is often realized via deploying multiple unsplittable paths, which is an NP-hard optimization problem as well, both for minimizing the number of paths and for maximizing $k$-splittable flows, cf.~\cite{BaierKoehlerSkutella2005,flowdecomposition}.
Another popular option is to split the flows at the routers using hash functions; 
other major techniques are flow(let) caches and round-robin splitting, cf.~\cite{DBLP:journals/network/HeR08}.
Nonetheless, splitting flows along multiple paths can lead to packet reordering problems, which need to be handled by further techniques, see, e.g.,~\cite{Kandula2007}.

Many of the discussed flow migration works rely on linear programming formulations:
Even though their runtime is polynomial in theory, the timely migration of large networks with many intermediate states is currently problematic in practice~\cite{swan}. If the solution takes too long to compute, the to-be solved problem might no longer exist, a problem only made worse when resorting to (NP-hard) integer programming for unsplittable flows.
As such, some tradeoff has to be made between finding an optimal solution and one that can actually be deployed.

Orthogonal to the problem of consistent flow migration 
is the approach of scheduling flows beforehand, 
not changing their path assignments in the network
during the update. We refer to the recent works by Kandula \etal\cite{DBLP:conf/sigcomm/KandulaMSB14} and Perry \etal\cite{Perry:2014:FCZ:2740070.2626309} for examples. Game-theoretic approaches have also been considered, e.g.,~\cite{DBLP:journals/tcs/HoeferMRT11}. Lastly, the application of model checking 
does not cover bandwidth~restrictions~yet~\cite{mcclurg2015efficient,DBLP:conf/nsdi/ZhouJCCG15}.

\rev{
\subsection{Summary and Insights}
Congestion-aware consistency is stronger than loop-freedom~\cite{Forster2016Consistent}, however it can be seen as orthogonal to policy consistency.
Notwithstanding, most algorithms tag individual flows and perform 2-phase commits~\cite{abstractions}, thereby achieving some levels of policy consistency.
Overall, four algorithmic techniques are currently studied:
$1)$ If flows must remain on the paths defined by their old or new state, dependency graph approaches~\cite{dionysus} are popular, which can also capture dynamic network conditions~\cite{DBLP:conf/conext/GandhiRJ17}.
$2)$ If flows may be spread over the network, then insights from flow augmentation algorithms can be applied~\cite{SebastianBrandt2016,roger-infocom}.
$3)$ If non-polynomial runtime is acceptable, the problem of finding short congestion-aware schedules can be formulated as a MIP~\cite{icnp-jiaqi}.
$4)$ Modifying the packet headers via tagging can be omitted in many cases by carefully tailoring the update schedules~\cite{DBLP:journals/corr/AmiriDSW16,DBLP:conf/icalp/AmiriDSW18,DBLP:journals/corr/abs-1805-06315}, however possibly at the cost of some policy consistency.
In order to break deadlocks, the flows themselves may be rate-limited~\cite{dionysus} respectively oversubscribed~\cite{icnp-jiaqi}, or the background traffic is to be reduced~\cite{swan}.
A summary of all discussed algorithmic approaches can be found in Table~\ref{flow-algo-table}.
Regarding the complexity classification of flow migration, one can think of it being analogous to classic multi-commodity flow problems: discrete constraints are NP-hard, whereas continuous constraints permit polynomial runtime~\cite{roger-infocom}.
However, while the problem remains in NP if every flow may only be touched once~\cite{dionysus}, the complexity is unknown otherwise~\cite{DBLP:conf/nca/FoersterJSW16}.
All complexity results are summarized in Tables~\ref{hardness results-table} and~\ref{flow-hardness-table}.
}

\subsection{Open Problems}
\label{subsec:congestion-problems}

The classification of the complexity of flow migration 
still poses many questions, cf.~Table~\ref{flow-hardness-table}:
If every flow can only be moved once, then the migration (decision) problem is clearly in NP. However, what is the decision complexity if flows can be moved arbitrarily often, especially with intermediate paths? Is the ``longest'' fastest update schedule 
 for unsplittable flows: linear, polynomial or exponential? In other words, is the problem complete in NP, PSPACE, or EXPTIME?
Related questions are also open for flows of unit or integer size in general.

The problem of migrating splittable flows without memory limits and without intermediate paths is still not studied either: It seems as if the methods of~\cite{roger-infocom} and~\cite{Forster2016Power} also apply to this case, but a formal proof is missing.
Another open issue which researchers recently started to consider
concerns how to  
exploit traffic engineering flexibilities and helper rules to jointly optimize 
update scheduling and route selection~\cite{real-time}. \rev{It is also yet unclear how to integrate such helper rules into dependency graphs, beyond manually defining intermediate states that differ from old and new as in~\cite{DBLP:conf/hotnets/SinghGFFG17,DBLP:conf/sigcomm/SinghGFFG18}.}

Lastly, it would be interesting to compare the power and performance of the node-ordering approach introduced by Amiri \etal\cite{DBLP:journals/corr/AmiriDSW16}, \rev{further detailed in~\cite{DBLP:conf/icalp/AmiriDSW18,DBLP:journals/corr/abs-1805-06315}}, to using the 2-phase commit of Reitblatt \etal\cite{abstractions}.
\rev{Such a comparison could use involve the work of Zheng \etal\cite{DBLP:conf/icdcs/ZhengLTFSCW18,Zheng2017}, under relaxed consistency guarantees, see the next section.}

\section{Further Considerations in Network Updates}\label{sec:discussion}

\rev{We have so far assumed a ``logically-centralized''
perspective on the algorithmic network update problem,
and mainly focused on strong notions of consistency.
This is also the focus in most existing literature
on the topic.
However, there also exist interesting first work
on solutions trying to relax these assumptions,
by studying relaxed notions of consistency
and aspects of distributed control planes.
In the following, we summarize the most
important work.}

\subsection{Relaxing \rev{Consistency} Guarantees}\label{sec:orthogonal}

So far we studied network updates assuming that consistency in the respective model must be maintained, e.g., no forwarding loops must appear at any time.
\rev{
There are cases where consistency properties fundamentally cannot be guaranteed across an SDN network.
For example, Panda \emph{et al.}~\cite{cap-hotsdn13} noted that consistency (in terms of consistent application of some policies), availability and partition tolerance cannot be all guaranteed at the same time in an SDN network.
}
In situations where the consistency property cannot be maintained at all or the computation of consistent updates is not tractable, some works proposed to break consistency in a controlled manner.

A first approach in this direction consists in trying to minimize the time spent in an inconsistent state, with underlying protocols being able to correct the induced problems (e.g., dropped packets are re-transmitted), as done in  Google's \emph{B4} network~\cite{b4}, \rev{\cite{DBLP:conf/sigcomm/HongMAZABBJKLMP18}}.
\rev{This can be understood as a very relaxed form of consistency, eventual consistency~\cite{6133253,Forster2016Consistent,DBLP:journals/cacm/BailisG13}. 
}

\rev{For less relaxed guarantees, i.e., beyond eventual consistency, Mizrahi \emph{et al.} propose to} synchronize the clocks in the switches so that network updates can be performed simultaneously: With perfect clock synchronization, lossless communications and switch execution behavior, loop freedom could be maintained.  As the standard Network Time Protocol (NTP) does not have sufficient synchronization behavior, the Precision Time Protocol (PTP) was adapted to SDN environments in~\cite{DBLP:journals/ijnm/MizrahiM16,ISPCS14}, achieving microsecond accuracy in experiments. This obviously comes with additional message overhead for time synchronization in the whole network. An introduction and overview of so-called timed consistent updates is provided in~\cite{DBLP:journals/ton/MizrahiSM16}. 
Zheng \etal\cite{DBLP:conf/icdcs/ZhengLTFSCW18}, \cite{Zheng2017} study the use of timed consistent updates in order to prevent congestion in the context of flow migration, in combination with latency considerations~\cite{DBLP:conf/nca/Foerster18}, \cite[\S6]{IEEEexample:phdurl}.

Nonetheless, in some situations synchronized updates can be considered optimal: E.g., consider the case in Fig.~\ref{fig:move up-swan} where two unsplittable flows need to be swapped~\cite{DBLP:conf/ons/MizrahiM14}, with no alternative paths in the network available for the final links. Then, synchronizing the new flow paths can minimize the induced congestion~\cite{mizrahi2016time4}. 
Synchronized updates cannot guarantee packet consistency on their own, as packets that are currently en-route may still encounter changed forwarding rules at the next switch.
Time can also be used similarly to a 2-phase commit though, by analogously using timestamps in the packet header as tags during the update~\cite{jaq10}, with~\cite{jaq10} also showing an efficient implementation using timestamp-based TCAM ranges. Additional memory, as in the 2-phase commit approach of Reitblatt \etal\cite{abstractions}, will be used for this method, but packets only need to be tagged implicitly by including the timestamp (where often 1 bit suffices~\cite{SWFAN16,jaq10}). In~\cite{jaq8} some additional methods are discussed on how to guarantee packet consistency by temporarily storing traffic at the switches.

Despite all those advantages, the proposed clock synchronization approaches do not prevent unpredictable variations of command execution time on network switches~\cite{dionysus}, motivating the need for prediction-based scheduling methods~\cite{NOMS16,RFC7758}.
Even worse, failures have an intrinsic, unavoidable cost in this approach. If a switch fails to update at all, the network can stay in an inconsistent state until the controller is notified and takes appropriate actions (e.g., computing another update). The same risk of inconsistencies holds if controller-to-switch messages are delayed or lost. In contrast, techniques based on sequential approaches can verify the application of sent update commands one by one, possibly moving forward (to the next update) or back (if a command is not received or not yet applied) with no risk of incurring safety violations.

\subsection{\rev{Updates in Distributed Control Planes}}

\rev{
Emerging large-scale SDNs will need to rely on
scalable architectures and 
distributed control planes. 
Besides scalability, control planes need to be physically distributed to ensure
availability and fault-tolerance,
to improve load-balancing, and to reduce overheads. 
Distributed control planes can be organized differently,
e.g., be either partitioned vertically (e.g.,~\cite{elasticon,onos})
or horizontally (e.g.,~\cite{devoflow,kandoo,phemius2014disco}), where switches are typically sharded among
controllers (e.g., accounting for geographic location or latency), 
or where different controllers are in charge of
different flow spaces. The design of a distributed control
plane is a distributed systems problem and different designs
come at different tradeoffs~\cite{jukka-sdn}.

Which specific architecture is used also has implications on 
the network update problem.
In general, to ensure consistency in network updates, additional coordination 
among different controllers
may be required which comes with overhead:
to guarantee consistency of
network operation, actions performed on the data plane by
different controllers may need to be synchronized.
One option to this end are distributed data stores~\cite{DBLP:conf/ewsdn/BotelhoRKB13}, so that applications would ideally remain unaware of any inconsistency~\cite{DBLP:journals/pieee/KreutzRVRAU15}.
Especially for wide-area networks, such additional synchronization can add
substantial latency, where the notion of ``continuous consistency'' can be of use for parametrization in the application design of geo-replicated services~\cite{DBLP:journals/tocs/YuV02}.
On the other hand, strongly consistent network updates are unlikely
possible if the control plane itself is only weakly consistent.
\emph{STN}~\cite{infocom15} relies on a replicated state machine
to update networks, which provides strong consistency guarantees, namely linearizability:
thus, the distributed controller can emulate any existing network update
algorithm designed for a single controller.
However, it requires consensus. 
In contrast, 
\emph{Ravana}~\cite{DBLP:conf/sosr/KattaZFR15}'s consistency is based on a weaker notion
of ``observational indistinguishability'',
and \emph{Onos}~\cite{onos} and \emph{Net-Paxos}~\cite{netpaxos} rely on partial event ordering. 

Panda \etal\cite{cap-hotsdn13}
argue that linearisability is
often unnecessary for ensuring correct application
of most network policies as the investigated policies often have
simple correctness conditions. Motivated by this observation,
Sakic \etal\cite{7997164} aim to overcome the 
blocking process in strongly consistent distributed control planes, 
and propose an adaptive,  eventually consistent model.  
Levin \etal show in~\cite{levin2012logically} that distributed network functions
such as load-balancers can work around eventual consistency
and still deliver performance sufficient for production deployments.
Guo \etal\cite{DBLP:journals/cn/GuoSXDWHC14} further expand the work of Levin \etal by  reducing synchronization overhead.

Thanh \etal\cite{marco-dec} present ez-Segway, a decentralized mechanism to
consistently and quickly update the network state
while preventing forwarding anomalies (loops and blackholes)
and avoiding link congestion. In their design, the
centralized SDN controller only pre-computes information
needed by the switches during the update execution.
This information is distributed to the switches,
which use partial knowledge and direct message passing
to efficiently realize the update. This separation of
concerns has the benefit of improving update performance
as the communication and computation bottlenecks
at the controller are removed.

Related to the question of how to perform updates
in distributed control planes is the issue of how
to perform such updates \emph{in-band}: 
how to preserve connectivity between 
control and data plane if the updates performed by a remote controller 
also affects its own paths?
Guaranteeing that each switch is managed, at any
time, by at least one controller is challenging 
if control is in-band, 
and only recently, a first solution has been presented
by Canini \etal\cite{icdcs18selfstab}, based on self-stabilization
principles.
}

\rev{
\subsection{Summary and Insights}
Even though SDN comes with the promise of centralized control, the network itself remains a distributed system---a fact which is responsible for many of the difficulties encountered in the previous sections.
If one assumes perfect availability and/or partition tolerance, in clear contradiction to~\cite{cap-hotsdn13}, then providing consistency becomes much easier: updates can be assumed to be executed at perfectly synchronized points in time, without any faults.
Technical steps in this direction have already been performed by improving time synchronization protocols~\cite{DBLP:journals/ijnm/MizrahiM16} and implementing timestamp-based TCAMs~\cite{jaq10}, allowing for new algorithmic directions~\cite{DBLP:conf/icdcs/ZhengLTFSCW18,DBLP:journals/ton/MizrahiSM16,Zheng2017}.
Another hurdle is that at scale, the control is just \emph{logically} centralized. 
The underlying distributed control plane can be implemented in various ways, each with its own set of benefits and tradeoffs~\cite{jukka-sdn}; an additional challenge is introduced if control is in-band~\cite{icdcs18selfstab}.
}

\section{From Theory to Practice}\label{sec:practice}

As a complement to the previously-described theoretical and algorithmic results,
we now provide an overview on practical challenges to ensure consistent network
updates. We also describe how previous works tackled those challenges in order
to build automated systems that can automatically carry out consistent updates.
A brief overview indicating the current status is presented in Table~\ref{table:theorypractice}.

\begin{table*}[htbp]
\caption{Table indicating the current status of practical challenges to ensure consistent updates.}
\label{table:theorypractice}
\centering
\scriptsize
\begin{tabular}{|>{\centering}p{2.5cm}|>{\centering}p{6.9cm}|>{\centering}p{7.44cm}|}
\hline 
Challenge &  \underline{A}pproach / \underline{M}easurement & Status \tabularnewline
\hline  
\hline
\multirow{2}{*}{\parbox{2.4cm}{(\ref{one}) Ensuring basic communication}} & A: command line interface of devices & Available in (traditional) networks~\cite{chen-pacman-09,igp-updates}  \tabularnewline \cline{2-3} 
 &  A: SDN controller programs and monitors & Deployed via, e.g., OpenFlow~\cite{openflow} and Network Information Bases~\cite{onix} \tabularnewline
\hline \hline
\multirow{6}{*}{\parbox{2.4cm}{(\ref{two}) Applying operational sequences}} & M: coordinator messages lost respectively not applied by all devices & Measured on commodity hardware~\cite{dionysus,fib-ack} \tabularnewline \cline{2-3} 
& A: status-checking commands and protocols & Evaluated in testbed~\cite{chen-pacman-09} \tabularnewline \cline{2-3} 
& A: lower-level packet cloning mechanisms & Evaluated in simulations~\cite{igp-updates}, data set at \url{https://inl.info.ucl.ac.be/softwares} \tabularnewline \cline{2-3} 
& A: active probing packets &Evaluated in (small) testbed~\cite{monocle} \tabularnewline \cline{2-3} 
& A: acknowledgement-based protocols & Evaluated in small testbed~\cite{fib-ack} \tabularnewline 
\hline \hline
\multirow{4}{*}{\parbox{2.5cm}{(\ref{three}) Working around device limitations}} & M: flow table size and setup rate limits statistics gathering & Measured on a common OpenFlow implementation on a switch~\cite{devoflow} \tabularnewline \cline{2-3} 
& M: high rule installation latency & Measured on various types of (SDN) switches ~\cite{sosr15latency,high-fidel,oflops,dionysus}\tabularnewline \cline{2-3} 
& A: adapt schedules dynamically & Evaluated in testbed~\cite{dionysus} \tabularnewline \cline{2-3} 
& A: eliminate redundant updates & Evaluated in Mininet~\cite{covisor},  algorithms~\cite{dsn16}, \rev{SDN testbed~\cite{dsn16-new}}\tabularnewline \hline \hline
\multirow{4}{*}{\parbox{2.6cm}{(\ref{four}) Multiple control-plane conflicts}} & A: pro-actively specifying computation of final rules & Implementation of~\cite{pyretic} available at \url{http://www.frenetic-lang.org} \tabularnewline \cline{2-3} 
& A: implementing coordination and locking primitives
on switches & Implementation of~\cite{liron-ccr} available at \url{https://github.com/lironsc/of-sync-lib}\tabularnewline \cline{2-3} 
& A: reactively detecting and possibly resolving
conflicts & Algorithms~\cite{infocom15} \tabularnewline \cline{2-3} 
& A: meta-algorithms  &  General theory~\cite{coexistence-controlplanes-15} \tabularnewline \hline \hline
\multirow{2}{*}{\parbox{2.4cm}{(\ref{five})  Updating the control-plane}} & A: hypervisor maintains history  & Evaluated in Mininet~\cite{hotswap-hotsdn} \tabularnewline \cline{2-3} 
& A: explicit state transfer & Evaluated in Mininet~\cite{hotswap-sosr} \tabularnewline \hline \hline
\multirow{4}{*}{\parbox{2.4cm}{(\ref{six})  Events occurring during updates}} & M: impact of link failures (IGP) &  Evaluated in simulations~\cite{igp-updates-ton}, data set at \url{https://inl.info.ucl.ac.be/softwares} \tabularnewline \cline{2-3} 
& A: enforcing per-packet consistency against packet-tampering adversary & Evaluation in small testbed~\cite{huafoum} \tabularnewline \hline

\end{tabular}
\centering
\end{table*}

\subsection{Ensuring Basic Communication with Network Devices}\label{one}
Automated update systems classically rely on a logically-centralized
coordinator, which must interact with network devices to instruct them to
apply operations (in a given order).
Such a device-coordinator interaction requires a communication channel.
Update coordinators in traditional networks typically exploit the command line
interface of devices, as noted in, e.g.,~\cite{chen-pacman-09,igp-updates}.
For SDNs, the interaction is simplified by their very architecture, since
the coordinator is typically embodied by the SDN controller which must be
already able to program (e.g., through OpenFlow~\cite{openflow} or similar
protocols) and monitor (e.g., thanks to a Network Information Base~\cite{onix})
the controlled devices.

\subsection{Applying Operational Sequences, Step by Step}\label{two}
Both devices and the device-coordinator communication are not necessarily
reliable. For example, messages sent by the coordinator may be lost or not be
applied by all devices upon reception~\cite{dionysus,fib-ack}. Those
possibilities are typically taken into account in the computation of the update
sequence (see \S\ref{sec:taxo}). However, an effective update system must also
ensure that operations are actually applied as in the computed sequences, e.g.,
that all operations in one update step are actually executed on the switches
before sending operations in the next step.
To this end, a variety of strategies are applied in the literature, from
dedicated monitoring approaches (based on available network primitives like
status-checking commands and protocols~\cite{chen-pacman-09}, lower-level
packet cloning mechanisms~\cite{igp-updates},
or active probing packets~\cite{monocle}) to
acknowledgement-based protocols implemented by SDN devices~\cite{fib-ack}.

\subsection{Working Around Device Limitations}\label{three}
Applying carefully-computed operational sequences ensures update consistency but
not necessarily performance (e.g., speed), as the latter also depends on device
efficiency in executing operations. This aspect has been analyzed by several
works, especially focused on SDN updates which are more likely to be applied in
real-time (e.g., even to react to a failure).
It has been pointed out that current SDN device limitations impact update
performance in two ways.
First, SDN switches are not yet fast to change their packet-processing rules, as
highlighted by several measurement studies.
For example, in the Devoflow~\cite{devoflow} paper, the authors showed that the
rate of statistics gathering is limited by the size of the flow table and is
negatively impacted by the flow setup rate.
In 2015, He \etal\cite{sosr15latency} experimentally demonstrated the high rule installation latency of four different types of production SDN
switches.
This confirmed the results of independent studies~\cite{high-fidel,oflops}
providing a more in-depth look into switch performance across various vendors.
Second, rule installation time can highly vary over time, independently on any
switch, because it is a function of runtime factors like already-installed rules
and data-plane load.
The measurement campaign on real OpenFlow switches performed in
Dionysus~\cite{dionysus} indeed shows that rule installation delay can
vary from seconds to minutes.
Update systems are therefore engineered to mitigate the impact of those
limitations -- despite not avoiding per-rule update bottlenecks.
Prominently, Dionysus~\cite{dionysus} significantly reduces multi-switch update
latency by carefully scheduling operations according to dynamic switch
conditions. In addition, CoVisor~\cite{covisor} and~\cite{dsn16} minimize the
number of rule updates sent to switches through eliminating redundant updates.

\subsection{Avoiding Conflicts between Multiple Control-Planes}\label{four}
For availability, performance, and robustness, network control-planes are often
physically-distributed, even when logically centralized (as in the case of
SDNs with replicated controllers).
For updates of traditional networks, the control-plane distribution is
straightforwardly taken into account, since it is encompassed in the update
problem definition (see \S\ref{sec:history}).
In contrast, additional care must be applied to SDN networks with multiple
controllers: if several controllers try to update network devices at the same
time, one controller may override rules installed by another, impacting the
correctness of the update (both during and after the update itself).
This requires to solve potential conflicts between controllers, e.g., by
pro-actively specifying how the final rules have to be computed
(e.g.,~\cite{pyretic}), by implementing coordination and locking primitives
on switches (e.g.,~\cite{liron-ccr}), or by reactively detecting and possibly resolving
conflicts (e.g.,~\cite{infocom15}).
A generalization of the above setting consists in considering multiple
control-planes that may be either all distributed, all centralized, or  hybrid
(some distributed and some centralized). Potential conflicts and general
meta-algorithms to ensure consistent updates in those cases are described
in~\cite{coexistence-controlplanes-15}.

\subsection{Updating the Control-Plane}\label{five}
In traditional networks, data-plane changes can only be enforced by changing the
configuration of control-plane protocols (e.g., IGPs).
In contrast, the most studied case for SDN updates considers an unmodified
controller that has to change the packet-processing rules on network switches.
Nevertheless, a few works also considered the problem of entirely replacing the
SDN controller itself, e.g., upgrading it to a new version or replacing an old
controller with a newer one.
In particular, HotSwap~\cite{hotswap-hotsdn} describes an architecture that enables
the replacement of an SDN controller, by relying on a hypervisor
that maintains a history of network events. As an alternative, explicit state
transfer is used to design and implement the Morpheus controller platform
in~\cite{hotswap-sosr}.

\subsection{Dealing with Events Occurring during an Update}\label{six}
Operational sequences computed by update algorithms forcedly assume stable
network conditions. In practice, however, unpredictable events, like failures,
can modify the network behavior concurrently and independently from the
operations performed to update the network.
While concurrent events can be very unlikely (especially for fast updates), by
definition they cannot be prevented.
A few contributions assessed the impact of such unpredictable events on the
update safety.
For instance, the impact of link failures on SITN-based IGP reconfigurations is
experimentally evaluated in~\cite{igp-updates-ton}.
A more systematic approach is taken by the recent FOUM work~\cite{huafoum}, that
aims at guaranteeing per-packet consistency in the presence of an adversary able
to perform packet-tampering and packet-dropping attacks.

\section{Future Research Directions}\label{sec:openprob}

We have identified and already discussed (in
\S\ref{subsec:connectivity-problems},\ref{subsec:policy-problems} and
\ref{subsec:congestion-problems}) several open research questions to preserve
each of the consistency properties considered by prior work.
We now describe more general areas which we believe deserve more attention by
the research community in the future.

\subsection{Charting the Complexity Landscape}
Researchers have only started to understand the 
computational complexities underlying the network
update problem. 
In particular, many NP-hardness results have been
derived for general problem formulations
 for all
three of our consistency models: connectivity consistency, 
policy consistency, and \congestion~consistency.
So far, only
for a small number of specific models polynomial-time optimal
algorithms are known. Even less is known about approximation
algorithms. Hence, more research efforts would be needed to 
chart a clearer picture of the complexity landscape
for network update problems. 
We expect that some of these
insights will also have interesting implications on classic
optimization problems, such as combinatorial
reconfiguration~\cite{DBLP:conf/icalp/AmiriDSW18} problems.

\subsection{Tailoring Update Mechanisms
to Specific Networks}
\rev{Datacenter topologies are usually highly
connected and regular while wide-area networks
are more sparse and organically grown:
properties which may be exploitable towards
more efficient and faster network update
algorithms. 
Today, hardly anything is known about
mechanisms for such more specific 
graph classes. 
A conceptually similar challenge arises in the context of reconfigurable links, e.g., ``\textit{how to gracefully transition between two topologies}''~\cite{DBLP:conf/ancs/FoersterGS18}~or dynamic (e.g., wireless or cellular) networks.
Additionally, the latter environments are appropriate for network coding~\cite{DBLP:journals/ccr/SzaboNSGF15,7147730}, where to the best of our knowledge current work is oblivious to ensuring consistency for codes during updates, respectively how to facilitate network coding itself for, e.g., faster updates.
Tailoring existing algorithm
and designing new algorithms for network
topologies arising in practice hence
constitutes a theoretically and practically
very interesting area for future research.}

\subsection{Refining our Models for Specific Technologies}
While today's network models capture well the \emph{fundamental} constraints and tradeoffs in consistent network update problems, these models are 
relatively simple. Today, SDN is used and discussed in various contexts and for different reasons, from network virtualization in datacenters to network slicing in emerging 5G applications~\cite{DBLP:journals/cm/ZhouLCZ16,DBLP:conf/eucnc/SilvaMKSBTZB16,slicing},
\rev{but also in the context of smart grids~\cite{DBLP:conf/smartgridcomm/DorschKGHW14,DBLP:conf/IEEEicics/ZhangSLF13}, wireless (sensor) networks~\cite{DBLP:journals/icl/LuoTQ12,DBLP:conf/latincom/OliveiraMG14,DBLP:journals/wc/BernardosOSBCJZ14}, and (hybrid) enterprise and ISP networks~\cite{ethane,panopticon1,DBLP:conf/sosr/HongMBM16}.
}
These use cases come with different \emph{specific} requirements on the 
consistency (e.g., whether per-packet consistency is strictly needed) and performance (e.g., tolerable number of rounds), and also differ in terms of the available \emph{knobs} which can be used for the network update (e.g., helper rules may undesirable in the context of network slicing). 
Accordingly, our models need to be refined and tailored towards specific use cases.

\rev{
\subsection{New Update Problems raised by Stateful Applications}
Network update schemes also depend on the application,
e.g., traffic engineering applications can come with very different
requirements than load-balancing \cite{7314891}.
Especially the development of more advanced and complex applications
(e.g.,~\cite{corr-virt,nate-event}, \cite[\S 3]{mcclurg}) on top of SDN networks may create the need
to support new consistency properties, and to potentially preserve these new
properties during network updates.
This is particularly true for stateful applications, whose state is built or modified
by multiple flows.
As an illustration, consider a stateful security appliance (e.g., a firewall)
that checks if traffic is illegitimate on the basis of the
exchange of information between flow sources and destinations over multiple
interaction rounds -- e.g., checking that TCP connections are
never started by machines external to an enterprise network.
In this case, network updates must consistently move all the flows needed for
the security appliance to work correctly.
For example, consistent updates should ensure that flows from any source $s$ to
any destination $d$ cross the same security appliance traversed by flows from
$d$ to $s$; otherwise, the appliance can incorrectly interrupt connections during
the update (and potentially block the corresponding traffic for some time).
An interesting avenue for future research is represented by studying if and how constraints related to stateful applications change the complexity of update
problems, as well as by developing algorithms and strategies to efficiently deal
with such additional constraints.

}

\rev{
\subsection{Update Frequency}
Another research direction is with respect to the frequency of network updates. 
Regarding inter-datacenter networks, \emph{SWAN}~\cite{swan} proposes to update the network every few minutes; a choice heavily influenced by the computation time needed. As such response times are not acceptable for, e.g., applications requiring interactive traffic, part of the network resources cannot be optimized on the fly.
Similarly, in a smart/micro grid environment, security concerns demand policy consistency, but at the same time, update speed is critical for power grid applications after network failures. Hence, Jin \etal\cite{DBLP:journals/tsg/JinLHCWSL17} employ the fast 2-phase commit protocol~\cite{abstractions}, which enforces policy consistency, but not, e.g., capacity consistency.
Security policy concerns also prevail in enterprise networks, where dynamic updates are required to facilitate users' changing devices as part of BYOD (Bring Your Own Device) or workplace timetables~\cite{allied}.
It would be interesting to see how computation (and deploy) times can be significantly improved for more complex consistency properties.
One option could be to employ machine learning, as already used to obtain routing configurations~\cite{DBLP:conf/hotnets/ValadarskySST17}.
On the other hand, if the updates are not time-critical but just augment the system's behavior, e.g., by improving throughput, the tradeoff between update frequency and performance would be of further research interest. 
Fundamental groundwork on this perspective has already been presented by Destounis \etal\cite{ParisDestounisMaggiEtAl2016}, but, as the authors point out, ``\textit{the network updates problem is orthogonal and complementary}''~\cite{ParisDestounisMaggiEtAl2016} to this problem angle.
Another dimension to explore is how to reduce update
frequency by improving in-band mechanisms~\cite{devoflow}.

}

\subsection{Dealing with Distributed Control Planes}
We believe that researchers have only started to understand the design and
implication of SDN control planes which are distributed not only vertically but
also horizontally~\cite{jukka-sdn,disco-up,DBLP:conf/sosr/NguyenCC17}.
\rev{
The research work developed so far mainly focuses on how to ensure that
concurrent update operations performed by different distributed controllers
reach a consistent state.
Depending on the specific setting, many more problems have to be solved.
For example, if each distributed controller controls a different subset of
devices, it may only change subpaths traversed by some traffic: How to perform
updates that are consistent network-wide, in this case?
Introducing synchronization between controllers might be a building block
towards the final solution, i.e., for the distributed controllers to agree on
which operation to perform when.
Which kind of synchronization is needed in this case? Are there lightweight
forms of synchronization that preserve some consistency properties with low
impact on control overhead and update speed?
}

\rev{
\subsection{Supporting In-Band Updates}
Challenges of guaranteeing consistency throughout network updates are further
exacerbated if the controllers have in-band control over the data plane devices,
since they are also affected by the effects of update operations. 
As an example of additional constraints to be taken into account, the
distributed controllers may lose connectivity to some switches that still need
to be updated, in the case of in-band connectivity between the controller and
the devices. How does this constraint affect the type and amount of update
scenarios that can be carried out by different strategies (e.g., ordering-based
algorithms)? Are there practical workarounds to ensure that controllers always
have connectivity with the controlled switches, irrespectively of the update
ordering and uncontrollable events like possible packet losses?
}

\section{Conclusion}\label{sec:conclusion}

The purpose of this survey was to provide researchers
active in or interested in the field of network update problems
(and in particular Software-Defined Networks)
with an overview of the state-of-the-art.
\rev{To this end, besides summarizing and tabularizing current results, we also presented a classification of consistent network updates in form of a taxonomy.
As such, the multitude of different} models, techniques, impossibility results, and practical challenges
\rev{are put into context and also allow direct comparisons.
Furthermore, we identified and listed many at present open technical and algorithmic problems, but also pointed out currently overlooked gaps in the framework of consistent network updates.
Additionally, we presented a historical perspective, showcasing the possibilities in traditional networks, which can be of interest to operators investigating the migration to Software-Defined Networking.
Subsequently, we} discussed the fundamental new challenges introduced
in Software-Defined Networks, also relating them to classic
graph-theoretic optimization problems. 
\rev{
These new challenges open a wide landscape of possibilities for future research, ranging from the integration of new technologies, adapting to particular network types, or the rise of complex (stateful) applications on top of current deployments, to list just a few.
}

\section*{Acknowledgements}

For inputs and feedback, we would like to thank 
Marco Canini, Nate Foster, Dejan Kosti{\'c}, 
Ratul Mahajan, Roger Wattenhofer,
and Jiaqi Zheng, \rev{as well as the anonymous reviewers of this article.}


{\bibliographystyle{IEEEtran}
\bibliography{literature}
}

\begin{IEEEbiography}[{\includegraphics[width=1in,height=1.25in,clip,keepaspectratio]{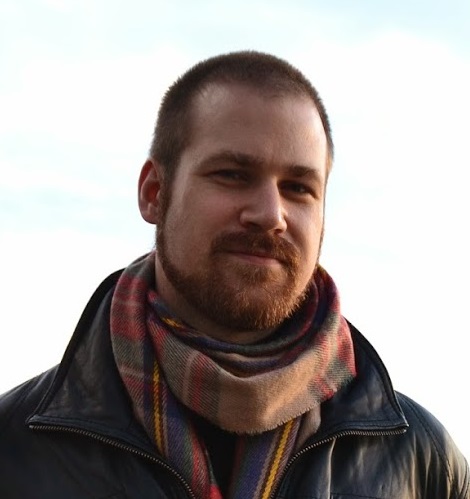}}]{Klaus-Tycho Foerster}
is a Postdoctoral Researcher at the Faculty of Computer Science at the University of Vienna, Austria since 2018. 
He received his Diplomas in Mathematics (2007) \& Computer Science (2011) from Braunschweig University of Technology, Germany, and his PhD degree
(2016) from ETH Zurich, Switzerland, advised by Roger Wattenhofer. 
He spent autumn 2016 as a Visiting Researcher at Microsoft Research Redmond with Ratul Mahajan, joining Aalborg University, Denmark as a Postdoctoral Researcher with Stefan Schmid in 2017.
His research interests revolve around algorithms and complexity in the areas of networking and distributed computing.
\end{IEEEbiography}
\vspace{-0.99cm}
\begin{IEEEbiography}[{\includegraphics[width=1in,height=1.25in,clip,keepaspectratio]{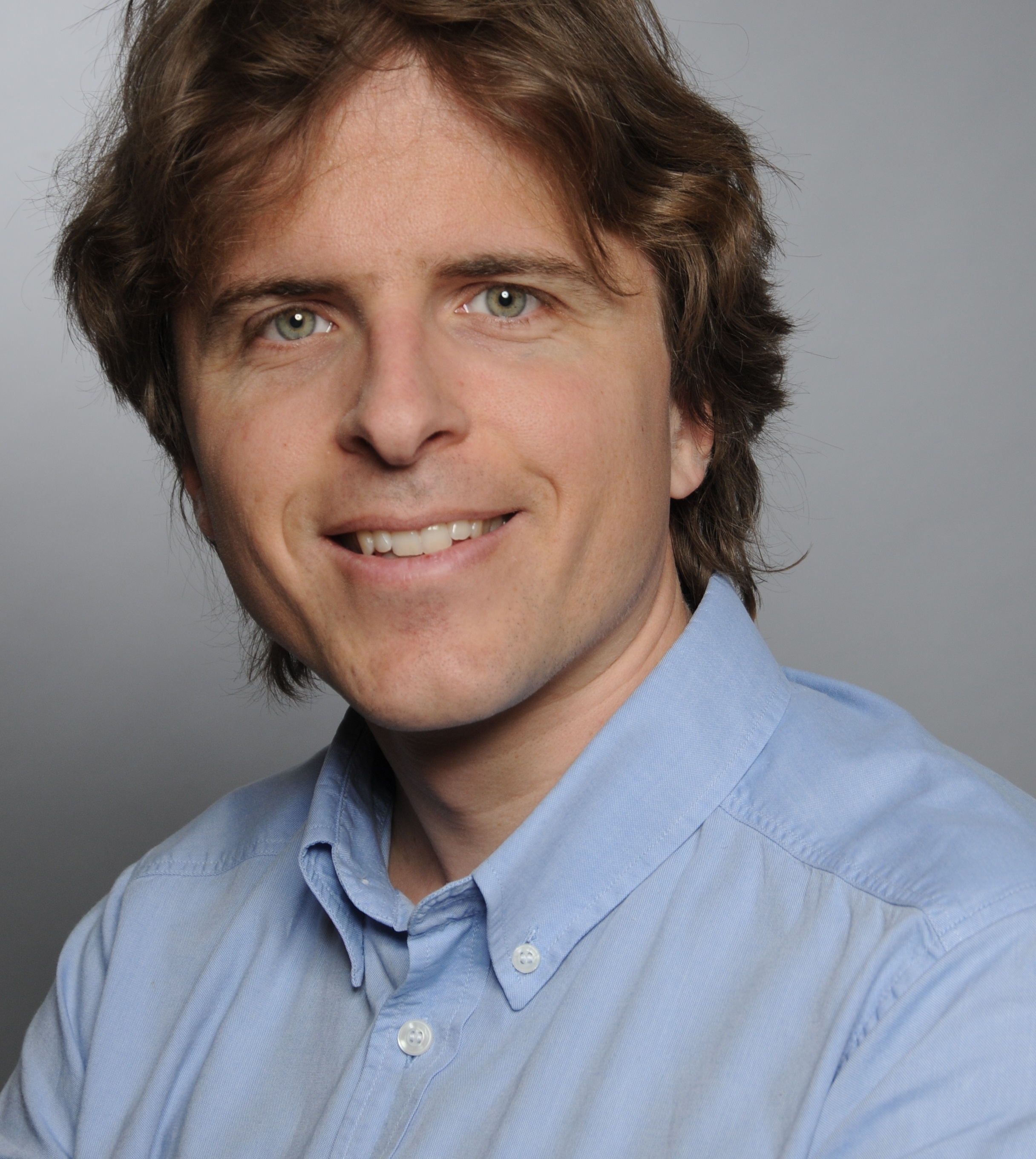}}]{Stefan Schmid}
is Professor
at the Faculty of Computer Science at the University of Vienna, Austria.  
He received his MSc (2004) and PhD degrees
(2008) from ETH Zurich, Switzerland. In 2009, Stefan Schmid 
was a postdoc at TU Munich and the University of Paderborn,
between 2009 and 2015, 
a senior research scientist at the Telekom Innovations Laboratories (T-Labs) 
in Berlin, Germany, and from the end of 2015 till early 2018, 
an Associate Professor at 
Aalborg University, Denmark. 
His research interests revolve around fundamental and algorithmic 
problems arising in networked and distributed systems. 
\end{IEEEbiography}
\vspace{-0.99cm}
\begin{IEEEbiography}[{\includegraphics[width=1in,height=1.25in,clip,keepaspectratio]{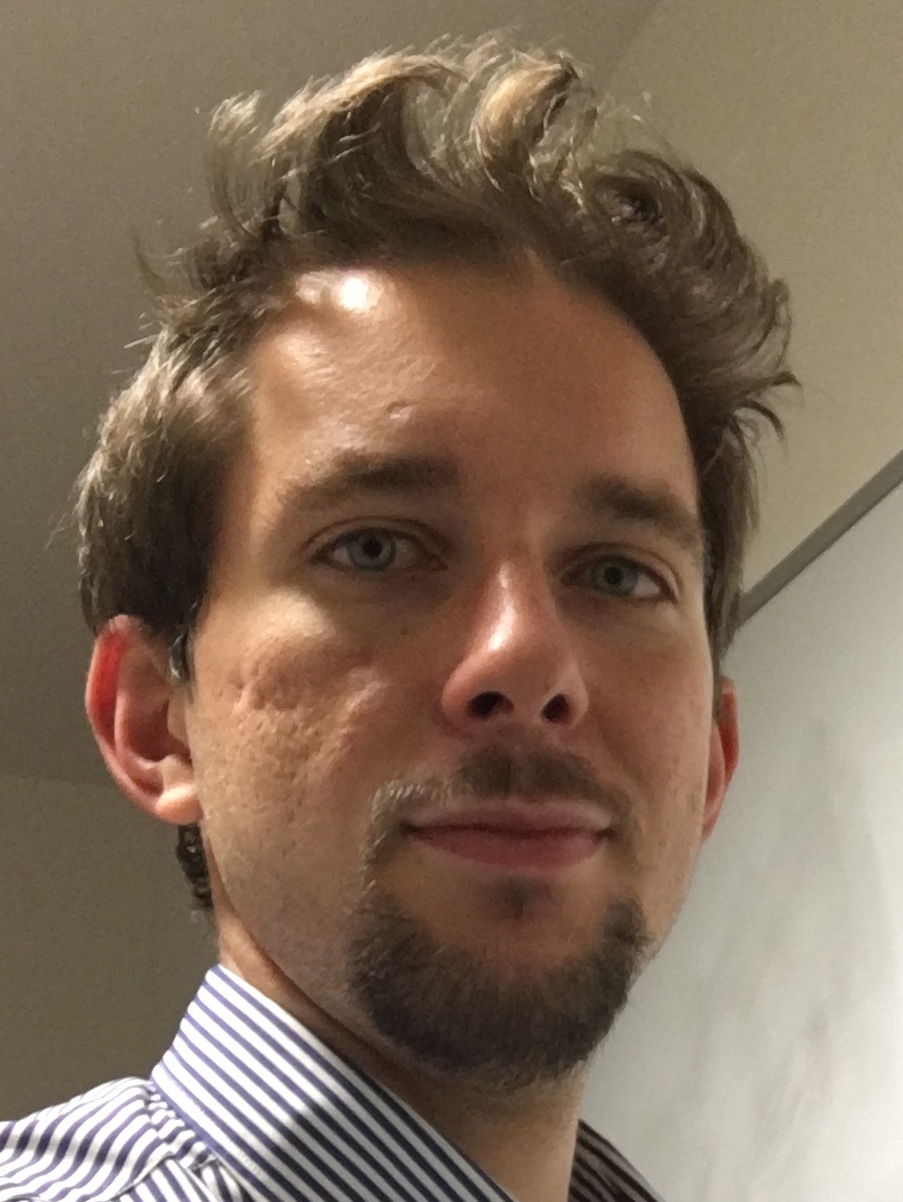}}]{Stefano Vissicchio}
is a Lecturer at University College London.
He obtained his Master degree from the Roma Tre University in 2008, and his
Ph.D. degree in computer science from the same university in April 2012. Before
joining University College London, he has been postdoctoral researcher at
the Universit\'{e} catholique of Louvain.
His research interests span network management, routing theory, algorithms
and protocols, measurements, and network architectures.
Stefano has received several awards including the ACM SIGCOMM 2015 best paper award,
the ICNP 2013 best paper award, and two IETF/IRTF Applied Networking Research
Prizes.
\end{IEEEbiography}

\end{document}